\documentclass[twocolumn,tighten,usenames,dvipsnames,twocolappendix]{aastex63}
\usepackage[normalem]{ulem}
\usepackage{verbatim}
\usepackage{amssymb}
\usepackage{amsmath}
\usepackage{xspace}
\usepackage{graphicx}

\usepackage{amsmath,mathtools,mathrsfs}
\usepackage{hyperref}
\usepackage{afterpage}
\usepackage{booktabs}

\usepackage[encapsulated]{CJK}
\usepackage{ucs}
\usepackage[utf8x]{inputenc}

\def\uas{$\mu$as\xspace} 
\def\uv{$(u,v)$\xspace} 
\newcommand{\themis}{{\sc Themis}\xspace}

\newcommand{\C}{\texttt{C}\xspace}

\renewcommand{\L}{{\mathcal L}}

\newcommand{\Npx}{N_{\rm px}}
\newcommand{\fov}{\xspace{\rm FOV}\xspace}
\newcommand{\SNR}{\xspace{\rm S/N}\xspace}
\newcommand{\AIC}{{\rm AIC}}
\newcommand{\BIC}{{\rm BIC}}

\defcitealias{M87_PaperI}{Paper~I}
\defcitealias{M87_PaperII}{Paper~II}
\defcitealias{M87_PaperIII}{Paper~III}
\defcitealias{M87_PaperIV}{Paper~IV}
\defcitealias{M87_PaperV}{Paper~V}
\defcitealias{M87_PaperVI}{Paper~VI}

\shorttitle{Hybrid VLBI Imaging and Modeling with \themis}
\shortauthors{Broderick et al.}

\begin{document}
  
\title{Hybrid Very Long Baseline Interferometry Imaging and Modeling with \themis}


\correspondingauthor{Avery E. Broderick}
\email{abroderick@perimeterinstitute.ca}

\author[0000-0002-3351-760X]{Avery E. Broderick}
\affiliation{ Perimeter Institute for Theoretical Physics, 31 Caroline Street North, Waterloo, ON, N2L 2Y5, Canada}
\affiliation{ Department of Physics and Astronomy, University of Waterloo, 200 University Avenue West, Waterloo, ON, N2L 3G1, Canada}
\affiliation{ Waterloo Centre for Astrophysics, University of Waterloo, Waterloo, ON N2L 3G1 Canada}

\author[0000-0002-5278-9221]{Dominic W. Pesce}
\affiliation{Center for Astrophysics $|$ Harvard \& Smithsonian, 60 Garden Street, Cambridge, MA 02138, USA}
\affiliation{ Black Hole Initiative at Harvard University, 20 Garden Street, Cambridge, MA 02138, USA}

\author[0000-0003-3826-5648]{Paul Tiede}
\affiliation{ Perimeter Institute for Theoretical Physics, 31 Caroline Street North, Waterloo, ON, N2L 2Y5, Canada}
\affiliation{ Department of Physics and Astronomy, University of Waterloo, 200 University Avenue West, Waterloo, ON, N2L 3G1, Canada}
\affiliation{ Waterloo Centre for Astrophysics, University of Waterloo, Waterloo, ON N2L 3G1 Canada}

\author[0000-0001-9270-8812]{Hung-Yi Pu}
\affiliation{ Perimeter Institute for Theoretical Physics, 31 Caroline Street North, Waterloo, ON, N2L 2Y5, Canada}

\author[0000-0003-2492-1966]{Roman Gold}
\affiliation{CP3-Origins, University of Southern Denmark, Campusvej 55, DK-5230 Odense M, Denmark}
\affiliation{ Institut f\"ur Theoretische Physik, Goethe-Universit\"at Frankfurt, Max-von-Laue-Stra{\ss}e 1, D-60438 Frankfurt am Main, Germany}
\affiliation{ Perimeter Institute for Theoretical Physics, 31 Caroline Street North, Waterloo, ON, N2L 2Y5, Canada}

\begin{abstract}
  Generating images from very long baseline interferometric observations poses a difficult, and generally not unique, inversion problem.  This problem is simplified by the introduction of constraints, some generic (e.g., positivity of the intensity) and others motivated by physical considerations (e.g., smoothness, instrument resolution).  It is further complicated by the need to simultaneously address instrumental systematic uncertainties and sparse coverage in the $u$-$v$ plane.  We report a new Bayesian image reconstruction technique in the parameter estimation framework \themis that has been developed for the Event Horizon Telescope.  This has two key features: first, the full Bayesian treatment of the image reconstruction makes it possible to generate a full posterior for the images, permitting a rigorous and quantitative investigation into the statistical significance of image features.  Second, it is possible to seamlessly incorporate directly modeled features simultaneously with image reconstruction.  We demonstrate this second capability by incorporating a narrow, slashed ring in reconstructions of simulated M87 data in an attempt to detect and characterize the photon ring.  We show that it is possible to obtain high-fidelity photon ring sizes, enabling mass measurements with accuracies of 2\%-5\% that are essentially insensitive to astrophysical uncertainties, and creating opportunities for precision tests of general relativity.
\end{abstract}


\section{Introduction}
\label{sec:intro}

Generating images from radio wavelength very long baseline interferometric (VLBI) experiments presents a substantial computational and mathematical challenge.  Generally, the quantity that is directly observed is related to the spatial Fourier transform of the image, i.e., the complex visibilities, $V(u,v)$, defined in the spatial frequency plane $(u,v)$.  In principle, if $V$ were measured at all spatial frequencies (i.e., all points in the $u$-$v$ plane), and if it were well calibrated (i.e., the full complex $V$ were known), then image reconstruction would reduce to an inverse Fourier transform.  In practice neither of the above conditions is met.

First, even for experiments with the most densely sampled $u$-$v$ coverage possible, its finite extent precludes a unique image reconstruction.  That is, the image becomes unconstrained on angular scales smaller than ${\sim}\lambda/u_{\rm max}$ -- this is simply the interferometric manifestation of the diffraction limit.  However, more typical in VLBI is sparse $u$-$v$ coverage, consisting of densely sampled tracks and significant gaps in which no visibilities are measured.  As a result, the lack of uniqueness can become significant, extending even to intermediate- and large-scale features.  Second, and particularly within the context of millimeter-wavelength VLBI \citep[see, e.g.,][]{M87_PaperIII}, the full complex visibilities are typically not available.  Atmospheric phase delays induce potentially large station-specific phase errors.  Imperfect antenna and receiver performance results in station gain amplitude errors.  Both of these may be ameliorated by careful treatment of correlations among baselines, e.g., self-calibration, or via the use of closure quantities, e.g., closure phase and/or closure amplitudes, or some combination thereof.  Again, these complicate and potentially alter the uniqueness of possible image solutions.  Both of these are relevant for observations made by the Event Horizon Telescope (EHT), a global millimeter-VLBI array \citep[][hereafter Papers~I, II, and III, respectively]{M87_PaperI,M87_PaperII,M87_PaperIII}.

Nevertheless, a number of efficient and high-fidelity methods have been developed to produce image reconstructions from interferometric data.
These include the venerable CLEAN algorithm \citep{Hogbom_1974,Schwarz_1978,Clark_1980,Schwab_1984}, which works directly with the inverse Fourier transform of the calibrated visibility data (the so-called ``dirty image'').  CLEAN attempts to model the image structure as a sum of point sources, whose locations and intensities are determined by iterative subtraction of appropriately scaled and shifted versions of the point spread function (the so-called ``dirty beam'') from the dirty image.  This procedure is typically repeated with interleaving ``self-calibration'' steps that attempt to infer the station-specific gain and phase errors consistent with the modeled image.  The resulting collection of point sources is subsequently smoothed to the ostensible diffraction scale, i.e., the radio ``beam,'' for display purposes, though the goodness-of-fit is computed directly from the point sources. 

A number of imaging methods have also been developed that attempt to forward model the image directly, traditionally using ``maximum entropy'' methods \citep[e.g.,][]{Frieden_1972,Gull_1978,Narayan_1986}, and more recently with ``regularized maximum likelihood'' (RML) methods \citep[e.g.,][]{Chael_2016,Chael_2018,Akiyama_2017a,Akiyama_2017b}.  These algorithms have two main advantages over the traditional ``inverse modeling'' approach used by CLEAN: (1) they permit the imposition of general-purpose regularizers on the image structure, such as positivity, smoothness, or sparseness; and (2) they can fit directly to closure quantities rather than requiring complex visibilities.  As with CLEAN methods, RML modeling may be iterated with self-calibration steps, during which the station-based gain terms may also be subject to prior constraints.  Both CLEAN and RML methods have been used with considerable success to reconstruct images of M87 from the 2017 April EHT observations \citep[hereafter, Paper~IV]{M87_PaperIV}.

In principle, the forward-modeling approach used by RML imaging methods makes them amenable to posterior exploration within a Bayesian framework, generating not just a single optimal fit but the full range of images consistent with the data. Access to the posterior is highly desirable because it enables a quantification of the non-uniqueness of the image reconstruction; i.e., it permits an estimation of the uncertainty in the image in both a formal and practical sense.  Posterior inference can also be used to naturally assess choices of resolution and field of view (\fov) of the image, systematically addressing two key parameters often left to the experience of the end user.

Previous efforts to perform Bayesian posterior exploration within the context of radio interferometric imaging have often focused on algorithms that function for connected-element arrays with excellent \uv-coverage and large FOVs \citep[e.g.][]{Sutter_2014}, motivating the development and use of sparsity priors \citep[e.g.][]{Cai_2018a,Cai_2018b} or resorting to single-point statistics that bypass the computationally taxing sampling \citep[e.g.][]{Greiner_2016,Junklewitz_2016,Naghibzadeh_2018}.  Nearly all of these efforts construct likelihoods that assume perfectly calibrated complex visibility data subject only to Gaussian thermal noise, and those that do attempt to fit for calibration terms either do not pursue image reconstruction \citep[e.g.,][who fit parameterized source models]{Lochner_2015} or impose simplifying approximations about the structure of the posterior \citep[e.g.,][who use variational inference]{Arras_2019}.

In this paper we present an imaging framework implemented within \themis, a flexible and highly parallel model comparison and parameter estimation code developed for the EHT \citep{themis}.  \themis has already been used to compare geometric and general relativistic magnetohydrodynamic (GRMHD) models to EHT data obtained from the 2017 April observing campaign \citep[hereafter, Papers~V and VI, respectively]{M87_PaperV,M87_PaperVI}.  It also provides a powerful set of existing sampling, modeling, and calibration tools, including samplers that can find and explore multi-modal likelihoods, and the ability to trivially generate new models by combining existing models \citep[see][for details]{themis}.  

In this way image reconstruction is treated in exactly the same standardized form as any other form of model fitting.  By leveraging the ability of \themis to explore image posteriors, we gain access to direct estimates of the pixel uncertainties and their spatial correlations throughout the image.  Perhaps even more importantly, the implementation of a \themis-based image model permits the self-consistent introduction of additional parameterized model components to the image.  Such components can include those on scales smaller than the image pixel size or larger than the image \fov, enabling the search for and characterization of features that may be infeasible to image but which may nevertheless be constrained by the data in the Fourier domain.

Here we present a non-parametric image model and example image reconstructions with and without additional geometric model components.  In particular, we demonstrate the ability to faithfully reconstruct bright ring-like features within images from GRMHD simulations appropriate for M87, and thus provide both a validation of the general underlying physical picture and a precision measure of the central supermassive black hole mass in M87  that is less sensitive to astrophysical uncertainties than prior methods. We summarize key relevant features of \themis in Section \ref{sec:themis}, present the image model in Section \ref{sec:imaging}, provide the introduction and validation of additional geometric features in Sections \ref{sec:hybrid} and \ref{sec:photon_rings}, respectively, and collect conclusions in Section \ref{sec:conclusions}.

\section{Summary of relevant \themis properties} \label{sec:themis}
\themis is a Bayesian model comparison and feature extraction framework developed for the EHT \citep{themis}.  Its modular design enables rapid development in most facets of the model analysis procedure.  Two aspects are particularly relevant here: the ability to add any image-based models rapidly and implement new samplers. 

All work reported here makes use of the parallel-tempered, differential-evolution Markov chain Monte Carlo (MCMC) sampler.  This is an affine invariant, ensemble MCMC scheme, and thus optimizes the proposal distribution.  Multiple tempering levels are used to efficiently locate and explore multi-modal likelihood distributions.  This sampler has been demonstrated with prescribed/known likelihoods containing thousands of independent peaks \citep[see Section 6.1.2 of][]{themis}.  However, it is well known that this class of samplers can struggle when parameter sets exceed $\gtrsim100$ \citep{Huijser+2015}.  While this sampler performance is sufficient for current EHT applications, additional sampler development is warranted prior to applications with a larger separation between the ostensible image resolution and \fov. Currently, the outputs of this sampler are MCMC chains that contain the full posterior probabilities of parameters of a given model.

\themis already has a number of simple geometric models implemented.  These include asymmetric Gaussians and slashed rings, i.e., annuli with a linear brightness gradient across the ring \citep[see Section 4.2 of][]{themis}.  Methods for dynamically generating new image-based models from existing models extend these with minimal effort.  A key example is the ability to sum images to generate new models; note that this sum is performed after visibility construction, and thus does not impose resolution restrictions from one model on others.  Additional image-based models may be simply implemented, requiring only information about the interface and otherwise agnostic regarding the remainder of \themis.

\themis also incorporates the ability to mitigate key systematic uncertainties in EHT observations.  Chief among these are the uncertain station gains, associated with atmospheric propagation and absorption.  Phase delays are naturally addressed by the use of closure phases.  An efficient and accurate method for reconstructing station gain amplitudes,
even in large numbers, has been implemented \citep[see Section 5 of][]{themis}.  A simple gain model is effectively marginalized over after the construction of likelihoods, permitting \themis to reconstruct many gains without creating additional demands on the sampler.

Here we present analyses of various simulated data sets produced with $u$-$v$ coverages identical and systematic error budgets similar to the 2017 observations for M87. The analyses are performed using the visibility amplitudes and closure phases, with the standard signal-to-noise (\SNR) restrictions and adopting the minimally covariant closure phase set \citep{themis,Blackburn+2019}.
In all cases we simultaneously reconstruct the individual station gain amplitudes while fitting image models as described in \citet{themis} and demonstrated in \citetalias{M87_PaperVI}.

\section{Imaging with \themis} \label{sec:imaging}

The line between ``imaging'' and ``modeling'' is conceptually blurred, relying on a qualitative distinction between models that prescribe a specific class of features and those that do not.  We approach imaging with \themis as a model-fitting endeavor through the use of an effectively non-parametric image model.  In this case, we use the term ``non-parametric'' loosely; the model is formally parametric (see \autoref{sec:RasterModel} for specifics), but the form of this parameterization is chosen to be flexible and to avoid steering the model toward any particular image structure.

We begin with a description of the underlying image model, discuss how resolution and \fov are set in a systematic fashion, and then present validation examples.

\subsection{Bicubic raster model} \label{sec:RasterModel}

\begin{figure}
    \includegraphics[trim=40 70 40 70, clip,width=\columnwidth]{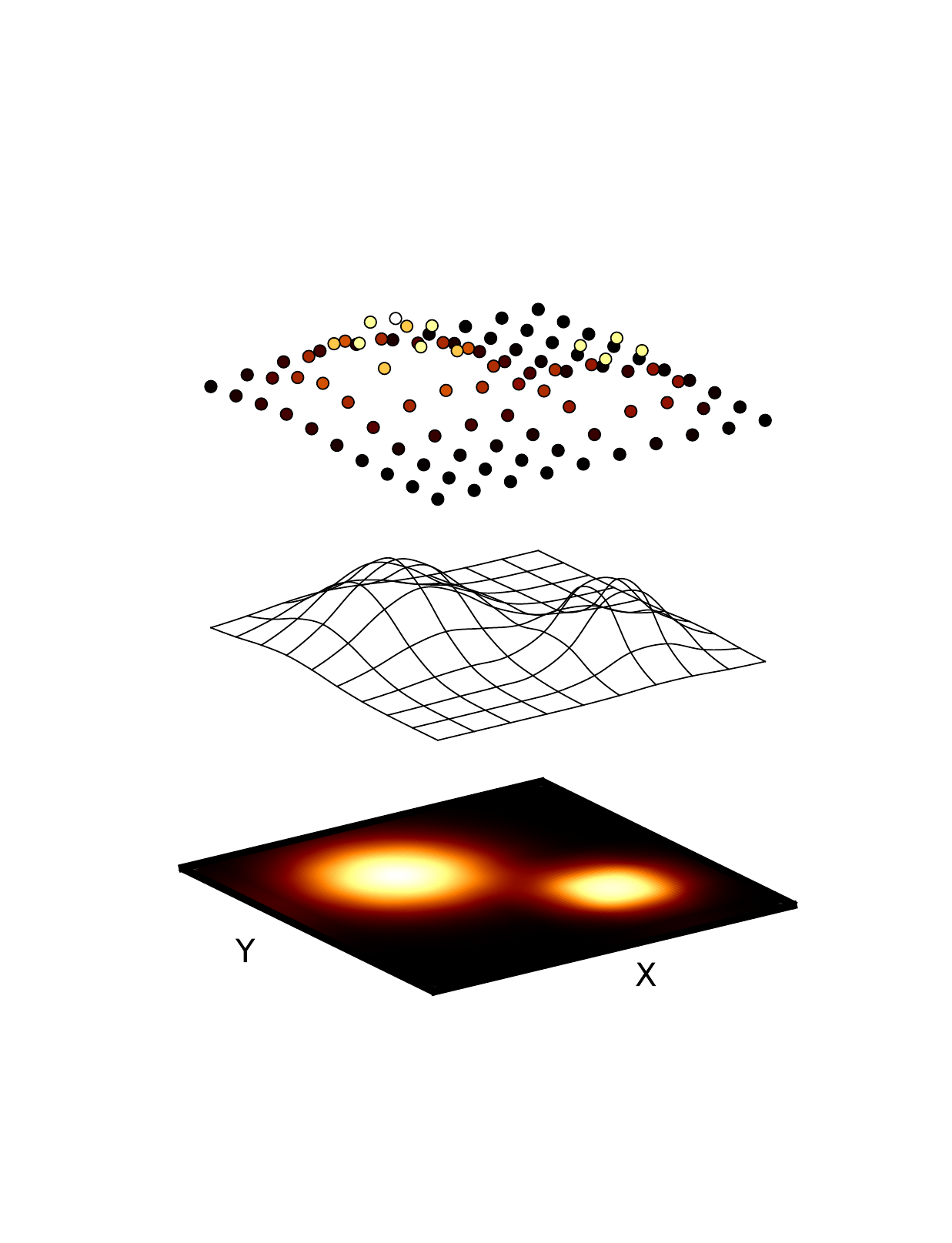}
    \caption{Cartoon of the image model described in \autoref{sec:imaging}.  Starting with a rasterized grid of control points (top), the intensities are interpolated using a bicubic spline (middle), resulting in a smooth representation of the image structure (bottom).}
    \label{fig:rastercartoon}
\end{figure}

The image model\footnote{This model is implemented within \themis using the \texttt{model\_image\_splined\_raster} class.} is generated in two steps: (1) amplitudes are selected for each of a number of ``control points,'' followed by (2) convolving those amplitudes with a bicubic smoothing kernel.

First, $\Npx$ control points are placed at the vertices of a uniform square grid contained within a pre-defined \fov, as shown in \autoref{fig:rastercartoon}.  These control points act as image pixels, the amplitudes of which are the primary model parameters; we enforce positivity at these control points by modeling the amplitudes logarithmically.  Given an \fov and a pixel number per side $N_{x,y}$ (such that $\Npx=N_x\times N_y$), we set the control point positions $(l_i,m_j)$ and intensities $I_{ij}$ to be
\begin{equation}
  l_i = s_x i,~~
  m_j = s_y j,~~
  I_{ij} = e^{p_{ij}}. \label{eqn:ControlPoints}
\end{equation}
where $s_{x,y}=\fov/(N_{x,y}-1)$.  We restrict the images considered in this paper to having $N_x = N_y = \sqrt{\Npx}$, such that the images are square and $s_x = s_y$.

Next, the grid of control points is convolved with a bicubic interpolation kernel to generate intensities that smoothly vary in all directions.  From the $I_{ij}$ values specified in \autoref{eqn:ControlPoints}, we generate the actual image intensity at every $(l,m)$ using
\begin{equation}
    I(l,m) =
    \sum_{i,j} w(l-l_i) w(m-m_j) I_{ij} , \label{eqn:ImageDomainConvolution}
\end{equation}
where
\begin{equation}
  w(x)
  =
  \begin{cases}
    0 & -2\ge x\\
    b\left[ -x^3 - 5x^2 - 8x - 4 \right] & -1\ge x>-2\\
    - (b+2) x^3 - (b+3) x^2 + 1 & 0\ge x>-1\\
    (b+2) x^3 - (b+3) x^2 + 1 & 1 \ge x > 0\\
    b\left[ x^3 - 5x^2 + 8x - 4 \right] & 2 \ge x > 1\\
    0 & x\ge2
  \end{cases}
  \label{eq:bciweight}
\end{equation}
is the 1D bicubic interpolation kernel and $b$ is a control parameter that affects the monotonicity of the interpolation.  For all of the images presented in this paper, we have set $b=-0.5$. We note that this bicubic interpolation can in principle result in negative values for $I$ at some $(l,m)$, despite the strictly positive nature of $I_{ij}$.  These departures from positivity are controlled by the value of $b$, and in practice we have found that setting $b=-0.5$ ensures that any negative excursions are typically very small.

In practice, we perform the convolution described by \autoref{eqn:ImageDomainConvolution} in the visibility domain.  That is, we set
\begin{equation}
  V(u,v) = \sum_{i,j} W(2\pi s_x u)W(2\pi s_y v) e^{2\pi i (l_i u + m_j v)} I_{ij} \,,
  \label{eq:V}
\end{equation}
in which 
\begin{multline}
  W(k) = - \frac{4}{k^3} \sin(k) \left[2b\cos(k)+(4b+3)\right]\\
  + \frac{12}{k^4} \left\{ b \left[ 1-\cos(2k) \right] + 2\left[1-\cos(k)\right] \right\},
  \label{eq:bcifour}
\end{multline}
is the Fourier transform of $w(x)$ (see \autoref{app:CubicInterpolation}).  For modest values of $N_x$, such as those presented here and relevant for the EHT, $V(u,v)$ is most rapidly computed via the direct sum in \autoref{eq:V}; fast Fourier transforms become advantageous only for $N_x\gtrsim10^2$.

In the absence of absolute phase information, as occurs here, the image is defined only up to an arbitrary translation.  However, the sparse nature of the available control points coupled with the assumed lack of flux outside of the prescribed \fov typically forces the image toward a single location, alleviating this concern. Thus, we make no attempt to fix the location of the image, e.g., by setting the center of light or imposing a prior on the spatial distribution of flux in the image. 

\subsection{Pixel size and FOV selection} \label{sec:resolution}

\begin{deluxetable*}{lccccc}
  \tablecaption{Dependence of the BIC (left entry) and AIC (right entry) on $N_x$ and \fov for a simulated data set designed to approximate the 2017 April 11 EHT observations of M87}\label{tab:Npxfov}
  \tablehead{
    \colhead{} &
    \multicolumn{5}{c}{FOV ($\mu$as)} \vspace{-0.15in}\\
    \colhead{} &
    \multicolumn{5}{c}{\rule{6.5in}{0.4pt}} \vspace{-0.100in}\\
    \colhead{$N_x$} &
    \colhead{40} &
    \colhead{60} &
    \colhead{80} &
    \colhead{100} &
    \colhead{120}
  }
  \startdata
  4  &    222 /    244  &    $(2.02$ / $ 2.04)\times10^{3}$   & $ (3.24$ / $ 3.24)\times10^{4}$  & $ (5.38$ / $ 5.38)\times10^{5}$  & $ (2.12$ / $ 2.12)\times10^{6}$ \\
  6  &  31.0 /  31.0  &  {\bf 0 / 0}  &    106 /    106  &    862 /    862  & $ (5.81$ / $ 5.81)\times10^{3}$ \\
  8  &    217 /    207  &    178 /    168  &    175 /    165  &    169 /    159  &    246 /    236 \\
  10  &    446 /    471  &    408 /    433  &    402 /    427  &    394 /    419  &    390 /    415 \\
  12  &    731 /    902  &    689 /    859  &    689 /    860  &    670 /    840  &    672 /    843 \\
  \enddata
  \tablecomments{Both information criteria are quoted relative to their minima, which occur for $N_x=6$, $\fov=60~\mu$as (in bold).}
\end{deluxetable*}

Within the context of VLBI, the smallest and largest recoverable image structures are approximately determined by the longest and shortest baselines in the array, respectively.  Information about the source structure on spatial scales that are larger than those probed by the shortest baselines, on scales that are smaller than those probed by the longest baselines, or indeed on any scales that are not directly measured by the array, must therefore be guided by the image model.  For our purposes, the relevant model hyperparameters are the control point separation (i.e., the pixel size) and the \fov, neither of which has a unique \textit{a priori} specification.  Other imaging methods have to contend with this same issue, but it is common practice to simply employ some rule of thumb when selecting these hyperparameters.

In our image modeling procedure, we would ideally pursue a strategy for selecting unique and data-driven values for both $N_x$ and \fov via Bayesian evidence maximization.  In practice, we approximate the Bayes factors using the Bayesian information criterion (BIC), which explicitly penalizes additional parameters:
\begin{equation}
\BIC = -2\mathcal{L}_{\rm max} + k \log( N_{\rm data} ) .
\end{equation}
Here, $\mathcal{L}_{\rm max}$ is the logarithm of the maximum likelihood, $k$ is the number of model parameters, and $N_{\rm data}$ is the number of data points.  We also make use of the Akaike information criterion (AIC), defined by
\begin{equation}
\AIC = -2\mathcal{L}_{\rm max} + 2k + \frac{2k(k+1)}{N_{\rm data}-k-1}.
\end{equation}

Both the BIC and the AIC are listed in \autoref{tab:Npxfov} for a simulated data set constructed to approximate the 2017 April EHT observations of M87.  Both information criteria converge on the same optimal values for \fov and $N_x$, yielding a natural choice for both hyperparameters.  We note that the selected \fov and $N_x$ values remain subject to our assumption of a raster geometry.  For example, we do not explore asymmetric or disconnected grids here.  The inferred resolution ($10~\mu$as) and extent of the emission region are very similar to those found by other forward-modeling approaches to image reconstruction applied to the same simulated data set \citetalias{M87_PaperIV}.  This implies a modest degree of superresolution, by roughly a factor of two, relative to the ostensible resolution of the EHT, approximately $20~\mu$as.

\subsection{Image reconstructions from synthetic data} \label{sec:ImageReconstructions}

\begin{figure*}
  \begin{center}
    \includegraphics[width=\textwidth]{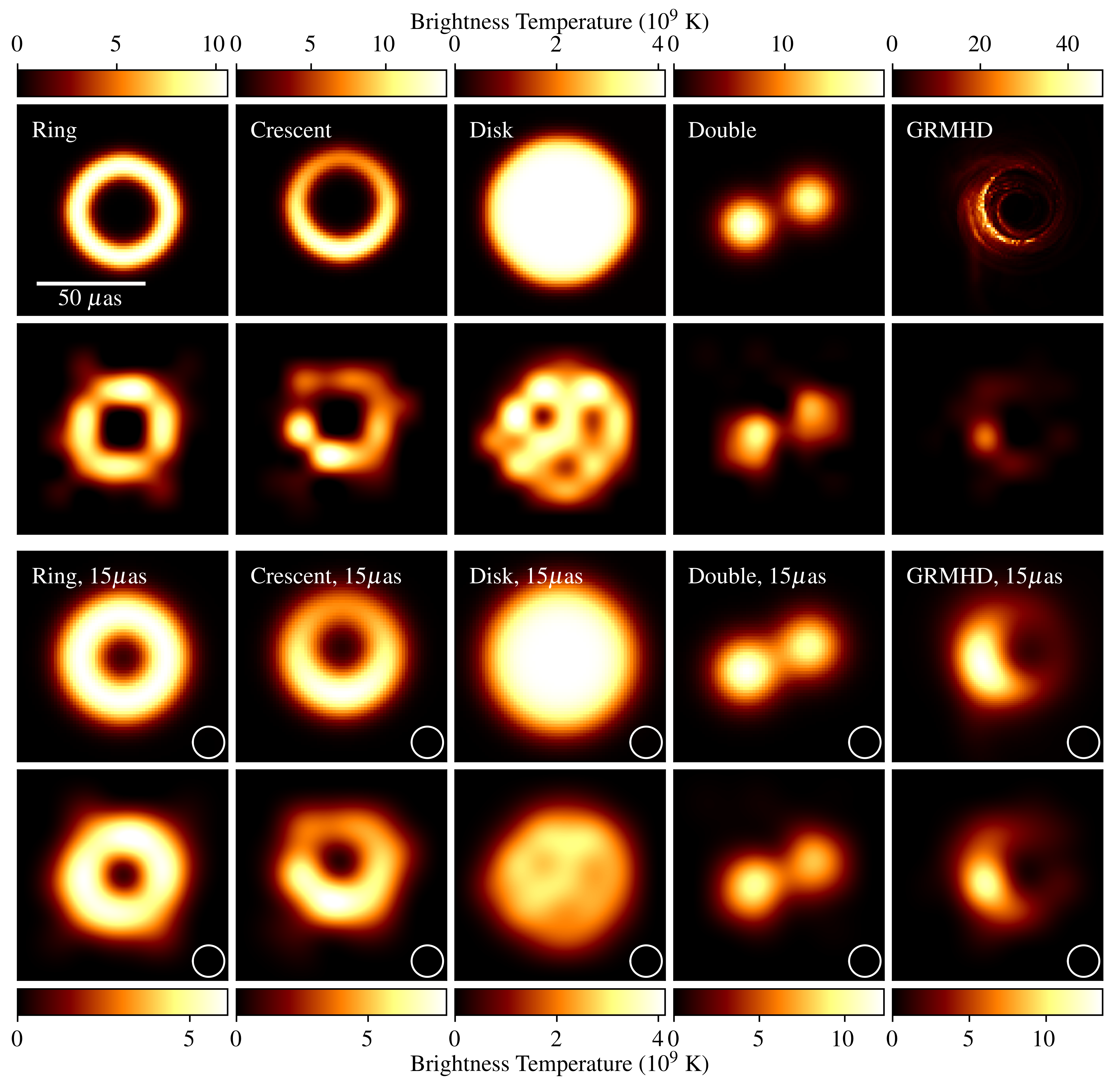}
  \end{center}
  \caption{Flux map reconstructions from the simulated data sets associated with the geometric models presented in \citetalias{M87_PaperIV}.  Each column corresponds to a single simulated data set, generated with the baseline coverage on April 11 and directly comparable to the reconstructions in Figure~10 of \citetalias{M87_PaperIV}.
  In all cases an FOV of $60~\mu$as and $N_x=6$ are chosen based on the resolution study for the GRMHD model (see \autoref{sec:resolution}).  The top row shows the underlying truth image.  The second row shows the hybrid image comprised of a bicubic raster model.  The third and fourth rows show the truth and reconstructed hybrid images convolved with a Gaussian beam with FWHM 15~$\mu$as, shown in the lower right corner; added in quadrature with the control point spacing this yields an effective resolution of roughly the ostensible EHT beam.  Each column has a fixed dynamic range, indicated by the associated color bars at the bottom.  Note that the total flux normalization is limited by the station gain amplitude uncertainties, which are known to roughly 20\%.}
  \label{fig:imagerecons}
\end{figure*}

In \autoref{fig:imagerecons} we present image reconstructions from a variety of simulated data sets with realistic $u$-$v$ coverage and noise for the 2017 April EHT observations of M87.  To facilitate comparison with other EHT imaging methods, we use the same simulated data sets for which reconstructions were presented in Figure~10 of \citetalias{M87_PaperIV}, which include realistic sources of systematic error, including large gain amplitude variations at a single station (Large Millimeter Telescope).  These images were selected to both characterize the ability of the imaging procedures to reconstruct and probe ringlike structures (e.g., Ring, Crescent, GRMHD) and to assess their ability to distinguish ringlike from non-ringlike structures (e.g., Disk and Double).

Prior to fitting, the data were pre-processed in the manner described in \citetalias{M87_PaperVI}; that is, we scan-averaged the visibility data and added a 1\% systematic error contribution (consistent with the magnitude of the non-closing errors measured in \citetalias{M87_PaperIII}) in quadrature to the thermal errors prior to constructing closure quantities.  We jointly fit to the closure phase and debiased visibility amplitude data products from both the high and low frequency bands, and we excluded all data points having a \SNR ratio below 2.  To accommodate potential structures that are resolved on all but the intrasite baselines \citepalias[e.g., those confined to Hawaii;][]{M87_PaperIV,M87_PaperVI}, all fits included a large-scale Gaussian component (see \autoref{sec:LargeScaleGaussian}).

In all cases we initialize the image model with a broad Gaussian emission structure, though subsequent exploration renders the details of the initial parameter values moot. All tests were run multiple times, and their MCMC chains have reached the well-mixed regime.  

\begin{figure*}
\begin{center}
\includegraphics[width=\columnwidth]{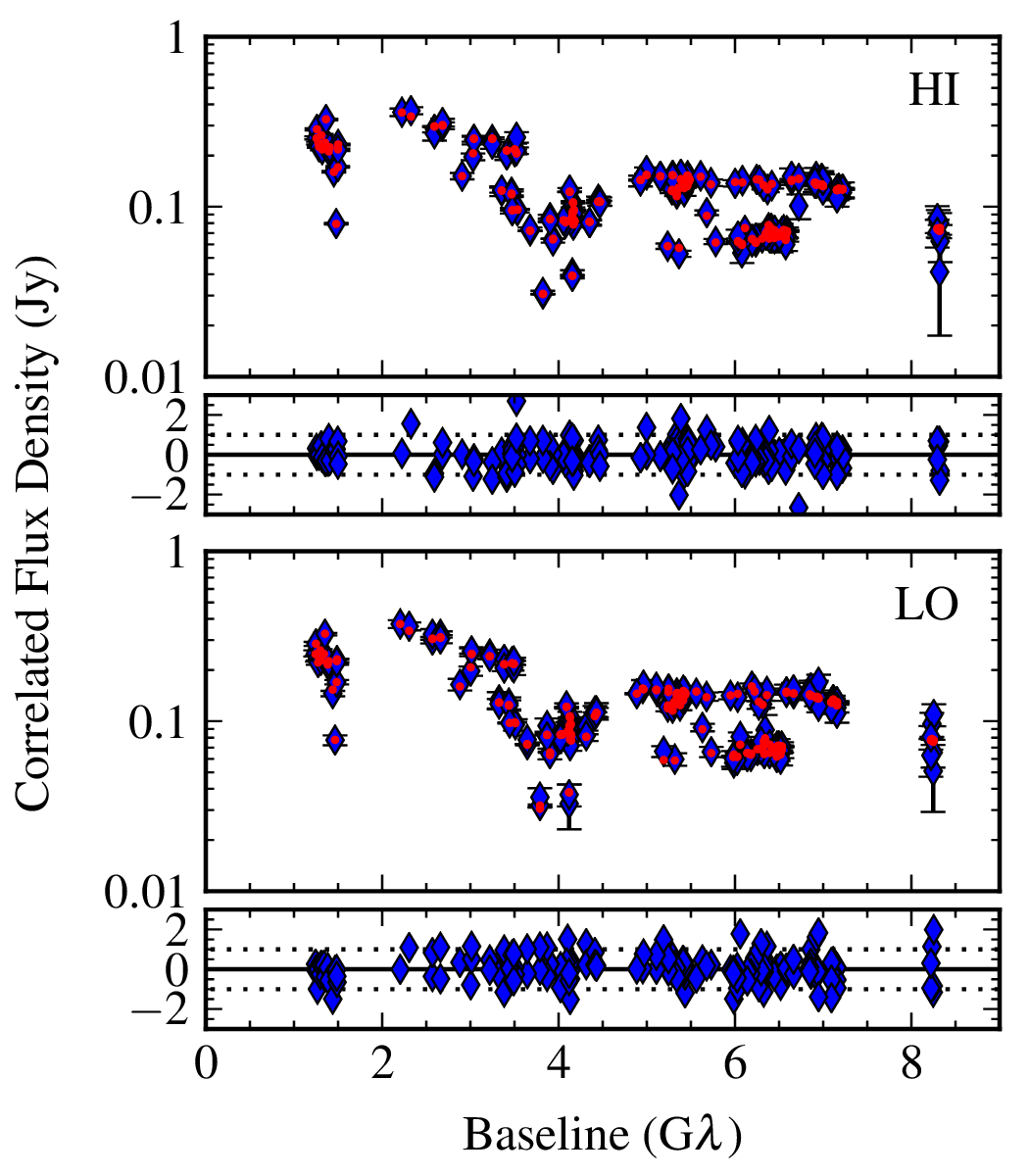}
\includegraphics[width=\columnwidth]{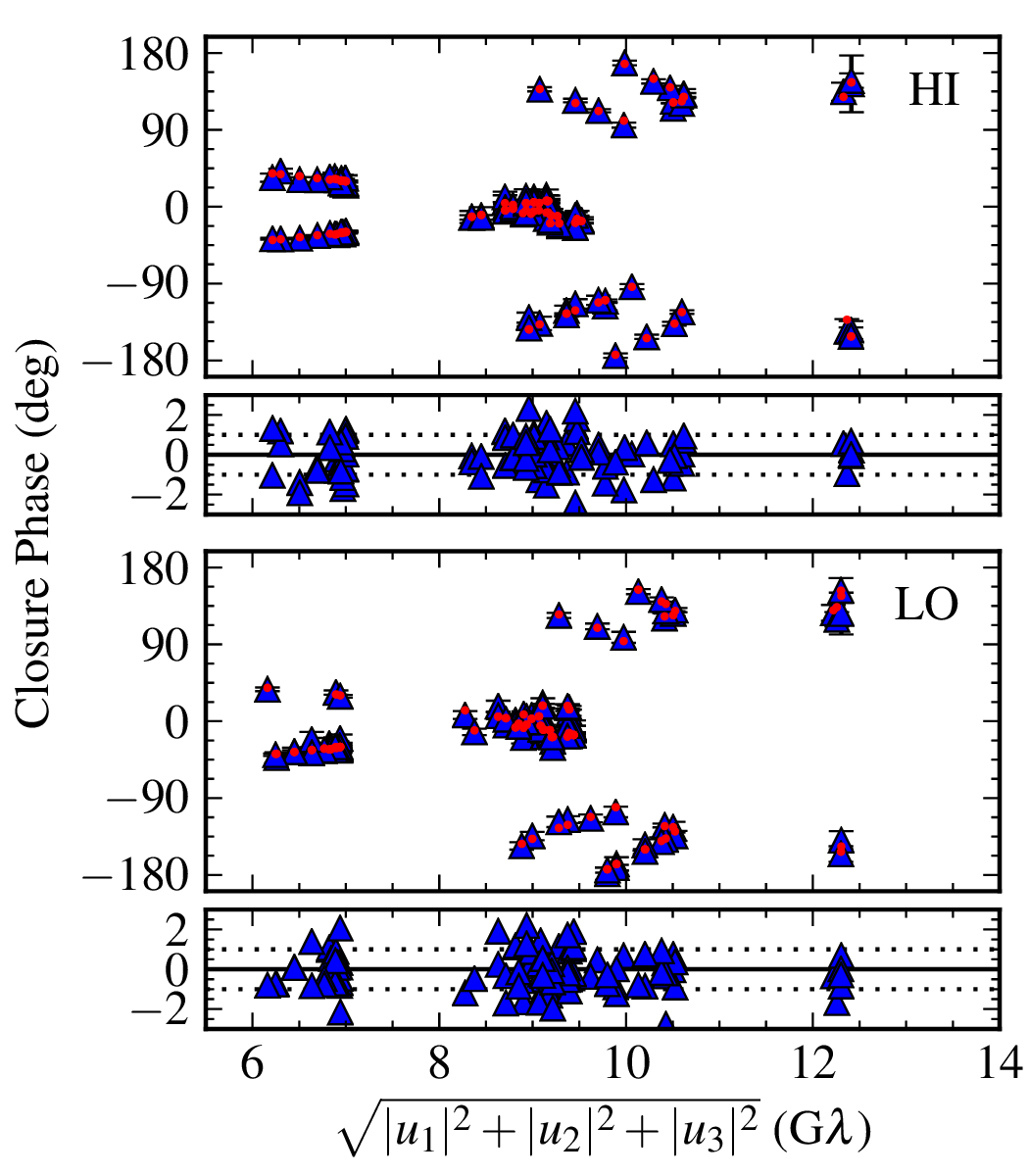}
\end{center}
\caption{Fit residuals for the GRMHD test shown in Figure~\ref{fig:imagerecons} for the high- and low-band visibility amplitudes (left, diamonds) and closure phases (right, triangles). In each set of panels the data are shown in blue and the maximum likelihood model is shown by red points.  Note that the reconstructed station gain amplitudes have been applied to the model values and not to the data.  Each comparison includes a subpanel that shows the normalized residuals directly underneath.}
\label{fig:M87residuals}
\end{figure*}

Good fits are found for all reconstructions, with the maximum likelihood fit achieving a $\chi^2$ per degree of freedom of $387/360$, $385/360$, $472/360$, $370/360$, and $330/360$ for the Ring, Crescent, Disk, Double, and GRMHD tests, respectively. An example set of residual plots is shown in \autoref{fig:M87residuals} for the GRMHD test, and we see no obvious structure or poorly fit sections of data.
To facilitate comparisons with other methods used for EHT image reconstruction, for each test we computed the equivalent blurring kernel defined and presented in Table~4 of \citetalias{M87_PaperIV}.  We find similar values to those reported there: $14.8~\mu$as, $14.6~\mu$as, $28.0~\mu$as, and $12.5~\mu$as, for the Ring, Crescent, Disk, and Double tests, respectively.

Despite the rectilinear nature of the control point gridding, a surprising variety of images is possible, as can be seen in \autoref{fig:imagerecons} (e.g., the reproduction of nearly circular Gaussian components in the Double test).  On the smallest angular scales accessible to the model -- i.e., the $10~\mu$as inter-pixel spacing -- we see imperfections in the maximum a posteriori reconstruction (e.g., knots on the Crescent, asymmetry in the components of the Double) associated with the limited \uv-coverage.  Smoothing the models with a $15~\mu$as FWHM Gaussian kernel mutes these small-scale structures and highlights the larger scales, for which the agreement between model and truth is more visually apparent (see the bottom two rows of \autoref{fig:imagerecons}).  However, we emphasize that even the apparent discrepancies in the unsmoothed model are statistically consistent with the truth, within the confidence levels claimed by the posterior.  That is, the locations and magnitudes of the various small ``artifacts'' are no more than expected given the data-driven uncertainties in each pixel's amplitude, and the myriad realizations of all possible small-scale discrepancies are fully captured by the posterior distribution (see \autoref{sec:ImagePosteriors}).

All of the image reconstructions shown in \autoref{fig:imagerecons} recover the scale and morphology of the truth images well.  This includes the sizes and separations between features, a brightness gradient when present and its absence when not present, and the overall image orientations.  Of particular relevance for detecting black hole shadows with the EHT is the ability to detect ring-like structures with high fidelity, that is, to reproduce rings when present in the underlying truth image but convincingly rule them out when they are not.

The overall flux normalization is a function of the station gain amplitude reconstruction, and thus subject to the associated priors.  This limits the flux reconstructions to a precision of roughly 20\% and is responsible for differences in the total brightness of some image reconstructions, e.g., the Disk test.

\subsection{Image posteriors} \label{sec:ImagePosteriors}

In practice, image parameter exploration as described in the previous sections is far more computationally expensive than the methods that have been used to date to generate images from EHT observations, which typically perform a constrained optimization of some comparison metric between image and data \citepalias[see][]{M87_PaperIV}.  This increased expense is a consequence of the general parameter estimation paradigm, which considers the best fit to be less important than the full set of compatible fits.  Thus, a key feature of image reconstruction within our modeling framework is the generation of a posterior distribution for all image parameters, which can be usefully conceptualized as an ``image posterior.''  The image posterior can be used to quantitatively address a variety of questions about a given image reconstruction.  Note that, despite the absence of absolute phase information, we do find that the image centroids are well fixed by the FOV combined with the modest number of pixels employed.  As such, we neglect the potential contributions to the statistics we present here from translations of the reconstructed image.

\begin{figure*}
  \begin{center}
    \includegraphics[width=\textwidth]{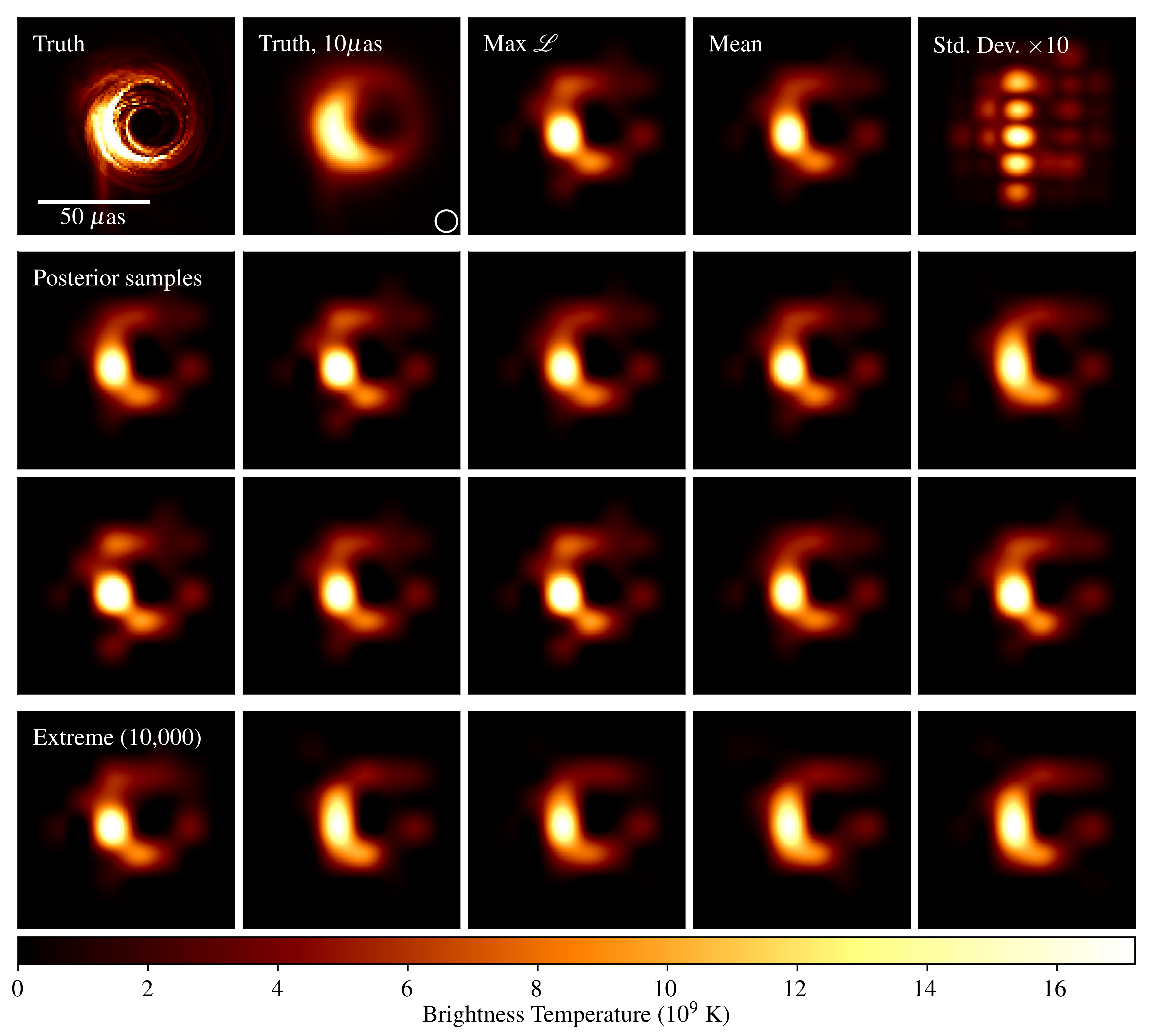}
  \end{center}
  \caption{Image reconstruction posterior for the 
  GRMHD test shown in Figure~\ref{fig:imagerecons}.  Top row: from left to right, the underlying truth image, truth image smoothed to the effective resolution of the reconstruction ($10~\mu$as), maximum likelihood reconstruction. In addition, the mean image, i.e., the image marginalized over the image posterior, and the standard deviation (multiplied by 10 for clarity) are shown.  Middle two rows: 10 random draws from the image posterior indicating the typical range in image reconstructions.  Bottom row: the five most extreme deviations in the image reconstruction from a sample of 10,000 images, showing the 3.5$\sigma$ outliers.}
  \label{fig:fitfamily}
\end{figure*}

A common question in radio interferometric imaging pertains to the reliability (or ``believability'') of specific image features.  In \autoref{fig:fitfamily} we show a selection of 15 reconstructed images taken from the MCMC chain for the GRMHD test image (fifth column of \autoref{fig:imagerecons}).  Ten of these images (shown in the middle two rows of the figure) are sampled directly from the posterior distribution, such that the magnitude and frequency of any variations seen across the images mirror those present in the image posterior (e.g., a feature seen in two of the 10 images indicates that $\sim$20\% of the image posterior space is consistent with that feature).  Additionally, the bottom row of \autoref{fig:fitfamily} shows the five most extreme\footnote{The degree of extremity here is quantified for any single image using the L2 norm of that image relative to the average image from the sample.} outlier images contained within $10^4$ elements randomly drawn from the MCMC chain; these images highlight image morphologies that live on the fringe of what the data will permit.

\begin{figure*}
  \begin{center}
    \includegraphics[width=\textwidth]{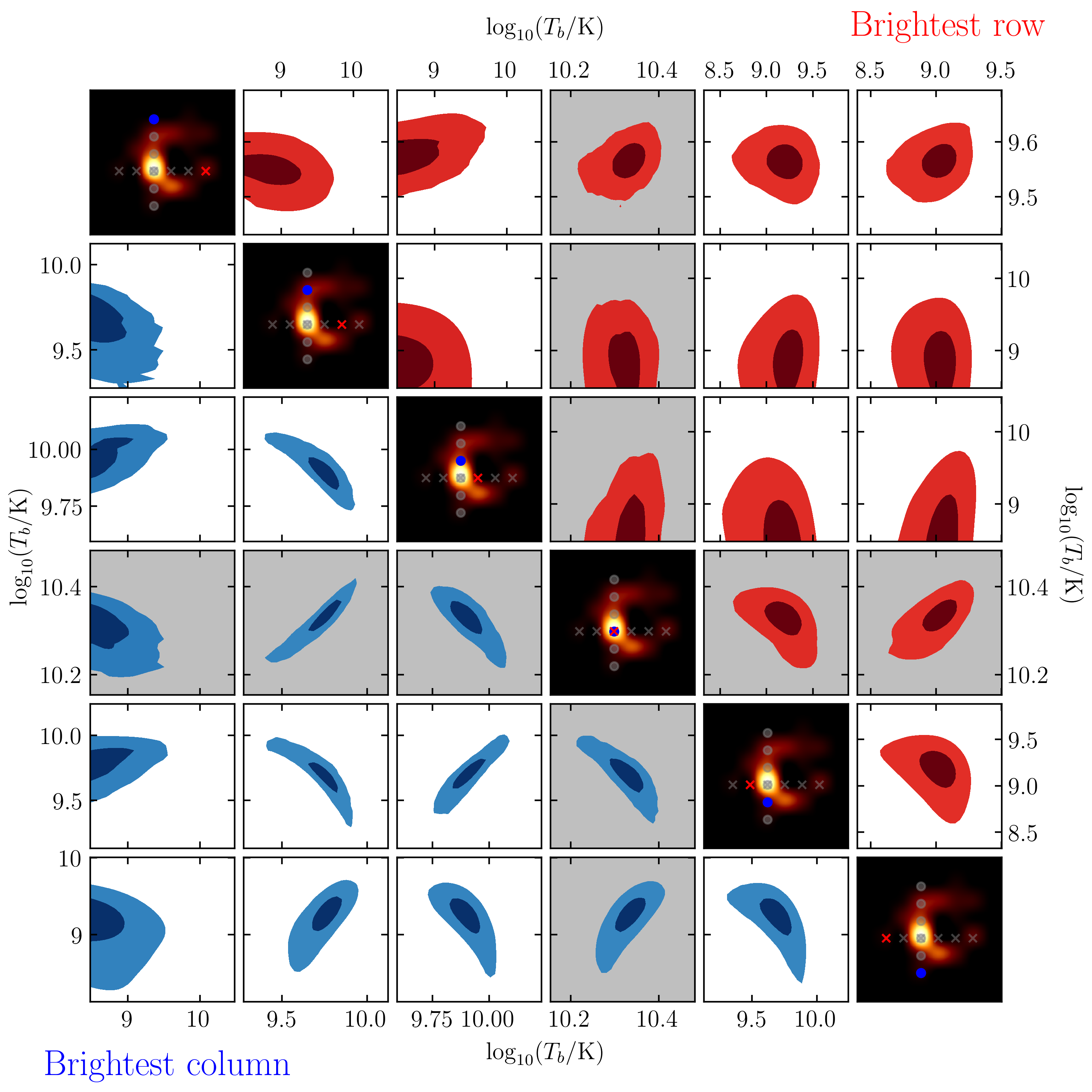}
  \end{center}
  \caption{Joint posteriors between a subset of the reconstructed pixels for the GRMHD test shown in Figure~\ref{fig:imagerecons}. Lower diagonal (blue) and upper diagonal (red) panels show the cross-correlations between the six pixels in the column and row containing the brightest pixel, respectively.  The positions of the pixel being correlated in each row/column are shown by the blue circles (lower diagonal) and red crosses (upper diagonal) within the images along the diagonal.  As an example, the rows/columns corresponding to the brightest pixel are shaded gray.  In all panels, contours enclose 50\% and 95\% of the posterior probability.  Note that where posteriors intersect with the low-brightness-temperature limits, they extend effectively to zero brightness temperature; this region has been excluded for clarity.}
  \label{fig:crosspx}
\end{figure*}

Single-point statistics computed from the image posterior can be used to produce ``typical'' images and to provide pixel-specific error budgets.  The top row of \autoref{fig:fitfamily} shows examples of such statistics in the form of the mean and standard deviation of the image posterior.  We find that the standard deviation map is highly inhomogeneous, varying substantially across the image domain; such behavior is both expected and frequently observed in radio interferometric imaging \citepalias[see, e.g., Fig.~17 of][]{M87_PaperIV}, despite often being presented as a single global rms value that is assumed to be shared by all pixels, but it is potentially critical for accurate image assessment.

Perhaps most striking is the clear ringing present within the standard deviation map, associated with the control point structure.  The origin of this ringing is elucidated by the pairwise marginal posterior distributions for pixels across the image, shown in \autoref{fig:crosspx} for a single row and column.  The nodes in the standard deviation map are associated with a set of sinusoidal oscillations with a wavelength of ${\sim}20~\mu$as, corresponding to the maximum spatial frequencies accessible to the 2017 EHT \uv-coverage.  For M87 in particular, limited north-south \uv-coverage (see Figure~12 in \citetalias{M87_PaperIII}) increases the uncertainty in the mode amplitude in that direction.  In contrast, we find little to no correlation between pixels in the east-west direction, where the 2017 EHT \uv-coverage was more complete for M87.  That is, the joint pixel flux distributions indicate both the presence of the expected resolution limit and the magnitude of the uncertainties this limit induces in the reconstructed image (roughly 10\% in this case).

With access to $n$-point statistics provided by the image posterior, arbitrarily sophisticated image interrogation procedures could be developed, such as metrics to assess how likely it is that the image contains a ring-like feature.  The ability to generate statistically rigorous posteriors is also critical for applications such as presented in \autoref{sec:hybrid}, where rigidly parameterized modeling is performed simultaneously with image reconstructions.

\section{Hybrid imaging with \themis}
\label{sec:hybrid}

Visibility-domain modeling of VLBI data can be a powerful tool for robust parameter estimation, but it assumes that the underlying source structure is well-described by the model under consideration.  This assumption is rarely satisfied for simple models when the data have high \SNR, at which point it becomes necessary to include additional and flexible model complexity to account for deviations between the desired model structure and the true source structure.  For a primary EHT science target, M87, the additional complexities have been associated with some combination of stochastic fluctuations in the emission region (as caused by, e.g., turbulence) and uncalibrated systematic uncertainties in the data.  The flexibility provided by imaging has played a critical role in constraining the nature of these deviations and in guiding the introduction of ad hoc model features \citepalias[e.g., the ``nuisance Gaussian components'' used alongside the geometric crescent models in][]{M87_PaperVI} to address them.

The implementation in the previous section of imaging within a modeling framework enables a novel path forward toward realizing the best elements of both approaches: (a) modeling known features and thus accessing substantial improvements in their parameter estimation, and (b) imaging the remainder of the source structure, yielding an otherwise agnostic view of the stochastic aspects of the image.  We refer to this simultaneous modeling and imaging approach as ``hybrid imaging.''\footnote{Note that the ``hybrid imaging'' presented in this paper is distinct from the ``hybrid mapping'' of, e.g., \citet{Baldwin_1978} and \citet{Pearson_1984}, which refers to an iterative procedure for reconstructing images without phase information.}

There are two key elements to successful hybrid imaging.  First, because a primary goal is the accurate reconstruction of the parameters of the model components, the ability to explore the posterior distribution for the full image/model combination is required.  Where simultaneous exploration is not possible -- e.g., when modeling separately from or iteratively with finding best-fit images -- the sampler will only have restricted access to the parameter space and the reconstructed posteriors will not necessarily be reliable.  Thus, the ability to construct full image priors as described in the previous section is critical.

Second, some ordering parameter in which the model and image components can be distinguished is highly desirable if not strictly necessary.  This ordering may be accomplished by enforcing a separation in angular scale, variability, polarization properties, wavelength, etc.  Where features in the image component can be reconstructed independently by either the model or image components, a degeneracy is induced between the two.  There are many problems for which these components are well separated in a natural way, and in this section we describe two such cases that are of particular relevance to EHT applications.

\subsection{Large-scale flux} \label{sec:LargeScaleGaussian}

The primary EHT targets exhibit structure over a wide range of angular scales, including on those much larger than the ${\sim}200$\,\uas \fov of the EHT images.  This large-scale image structure impacts only the shortest (intrasite, $<10$~km) baselines while having little effect on intermediate and long baselines \citepalias[see, e.g.,][]{M87_PaperIV,M87_PaperVI}.  Nevertheless, these short baselines provide important constraints on station gains and removing them would thus be undesirable.

\begin{figure*}
    \includegraphics[width=1.00\textwidth]{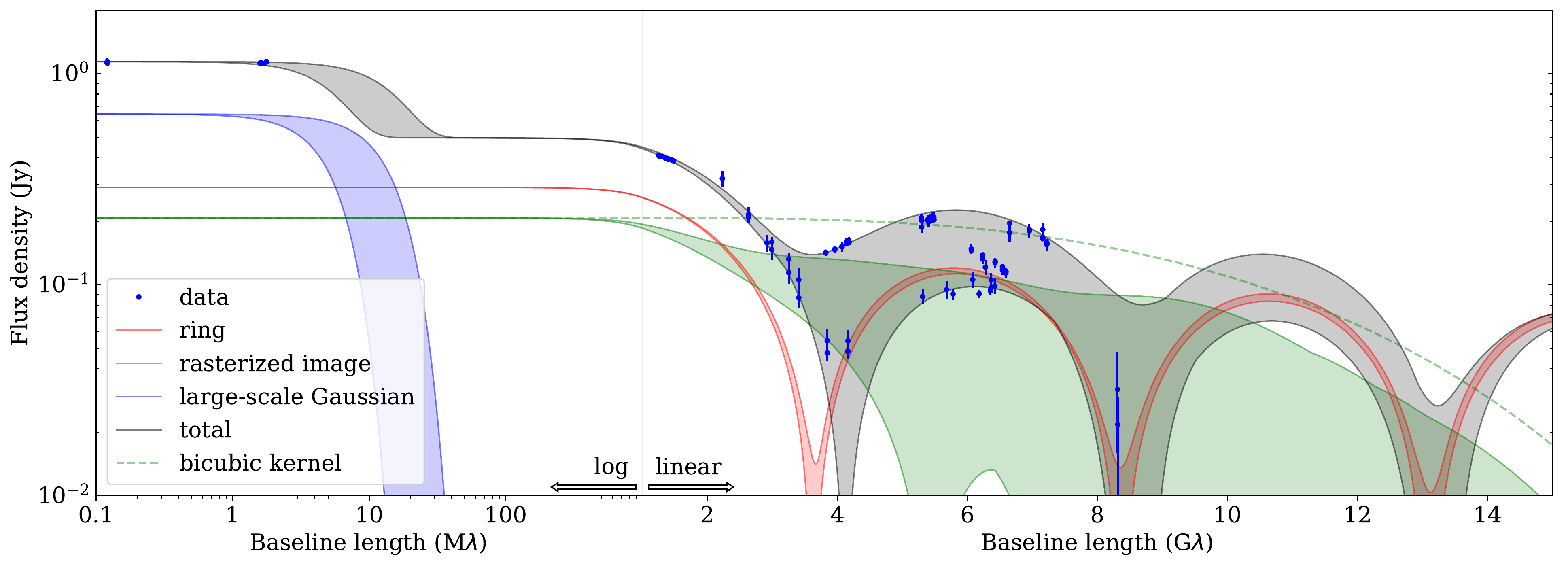}
    \caption{Visibility amplitude versus baseline length for the hybrid image model components considered in \autoref{sec:hybrid}.  The thin ring component is shown in red, the image component is shown in green, the large-scale Gaussian is shown in blue, and the combined model is shown in gray; the associated April 11 low-band synthetic data from the $a=0.5$, MAD GRMHD simulation image (see the first column of \autoref{fig:Jppcomp}) are plotted as blue points, after correcting for fitted gains.  The envelope corresponding to the kernel of the bicubic smoothing function (see \autoref{sec:RasterModel}) is plotted as a dashed green line.  To more easily illustrate the large scale separation, the horizontal axis switches from logarithmic spacing (left) to linear spacing (right) at a baseline length of 1\,G$\lambda$.  Interestingly, we see that the first minimum for the ring component falls at a smaller baseline length (corresponding to larger angular scale) than the first minimum for the combined model; this behavior can be understood in light of the structure seen in the top leftmost panel of \autoref{fig:Jppcomp}, which has the non-photon ring emission falling preferentially interior to the ring itself.}
    \label{fig:scale_separation}
\end{figure*}

To address the intrasite baseline flux excess, it has been found sufficient to include a large-scale ($\text{FWHM} \gtrsim 1$\,mas) Gaussian model component in the image generation and model comparison exercises. A Gaussian is chosen simply to have a structure-agnostic model component with a specifiable size scale. This component is naturally separated in spatial scale from the much more compact ($\text{FWHM} < 100$\,\uas) image features produced in the reconstructions (see \autoref{fig:scale_separation}).  We include such a Gaussian in all of the examples presented in this and the following section for this purpose.

\subsection{Photon rings}
Of more physical and algorithmic interest to the EHT is the ability to probe scales much smaller than the nominal array resolution.  Imaging algorithms currently in use by the EHT regularly achieve a modest degree of ``superresolution'' \citep[up to factors of $\sim$2--5;][]{Honma_2014, Chael_2016, Akiyama_2017b, Kuramochi_2018} -- i.e., the ability to recover image features that are smaller than some rule of thumb such as the Rayleigh criterion would nominally permit -- by imposing assumptions such as sparsity and smoothness in the reconstructed image.  Even more substantial resolution improvements are attainable by enforcing correspondingly stronger assumptions about the image structure, such as through the use of rigidly parameterized models, provided that the true image structure adheres to the imposed assumptions.  For instance, if the true image consisted of two point sources, then the optimal angular resolution would be achieved by fitting a double point source model to the data; our resulting ability to discern the source separation in this case could potentially be orders of magnitude better than the nominal array resolution.

A natural example of a parameterizable sub-beam source structure within the context of EHT observations is the ``photon ring.''  This actually consists of a concentric series of narrow, bright rings that originate from photons orbiting (``winding'') multiple times about the black hole.  Typically the largest of these rings, which generally contains the highest total flux, arises from photons that traverse behind the black hole only once prior to propagating toward Earth; we will call this feature the $n=1$ photon ring.  Higher-order rings are predicted by general relativity, though their flux decreases exponentially with winding number; the $n=\infty$ ring is the boundary of the black hole shadow \citep{Hilbert1917,Johnson_2019}.  The relationships between the widths, fluxes, and shapes of different rings provide additional tools for probing the underlying spacetime, though we leave a full discussion to future work; here, we focus on only a single bright, narrow ring.

The photon ring is not expected to be uniformly bright in general for many reasons: photons seen at different azimuthal locations on the ring have traversed different regions, and have propagated in potentially very different directions through a typically highly structured environment exhibiting relativistic motions and suffering large gravitational redshifts. In the case of M87, for which the inclination is ostensibly less than $20^\circ$, this photon ring is very nearly circular, deviating from circularity by less than 2\% \citep{Johnson_2019}.  We thus model the photon ring as a circular annulus with a linear brightness gradient \citep[see Section 4.2.5 of][]{themis}.  The key parameters are the total flux in the ring, outer radius of the ring, fractional width of the ring, and the magnitude and direction of a linear brightness gradient.  We enforce that the ring be narrow by restricting its width to be less than 5\% of the outer radius, corresponding to a width of roughly 1~$\mu$as, or about 10\% of the pixel resolution. This width prior also serves to enforce the component order, ensuring that the ring component and the image component of the model are not degenerate with one another.  The narrowness of the ring manifests in the Fourier domain as an essentially fixed peak-to-peak flux ratio; i.e., it sets the decay envelope for the Bessel function.  This envelope will in general not match that for the bicubic kernel (see \autoref{fig:scale_separation}), which is how the two components become decoupled. Note that the flux in the ring is allowed to vanish and hence we do not force such a ring into the image.

\section{Detecting and Constraining Photon Rings with the EHT} \label{sec:photon_rings}
In this section we describe a number of imaging experiments designed to assess and demonstrate the ability of existing and future EHT observations to detect and constrain photon rings.  We focus on several example simulations from the GRMHD library presented in \citetalias{M87_PaperV}, all of which contain a bright, narrow ringlike feature.  These simulations differ substantially in their other properties, however, including the brightness and structure of the extended emission, presence of additional ringlike and spiral features, asymmetric components, and dynamical structures. For GRMHD simulations appropriate for M87 we are able to faithfully reconstruct the properties of the bright ringlike feature, and recover its radius.

\subsection{Ring Reconstruction from GRMHD Simulations}
Five example GRMHD movie snapshots that are broadly consistent with the EHT 2017 observations of M87 were selected, listed in \autoref{tab:grmhdsims}.  All of these come from simulations that were deemed acceptable in \citetalias{M87_PaperV} and were resized to best match the EHT data and rescaled to a total flux of 0.6~Jy. Data were prepared with the $u$-$v$ coverage from the 2017 EHT observations of M87 from \citetalias{M87_PaperIII}.  The fitting procedure is identical to that described in \autoref{sec:ImageReconstructions}, but now with the inclusion of a ring model.  

\begin{figure*}
  \begin{center}
    \includegraphics[width=\textwidth]{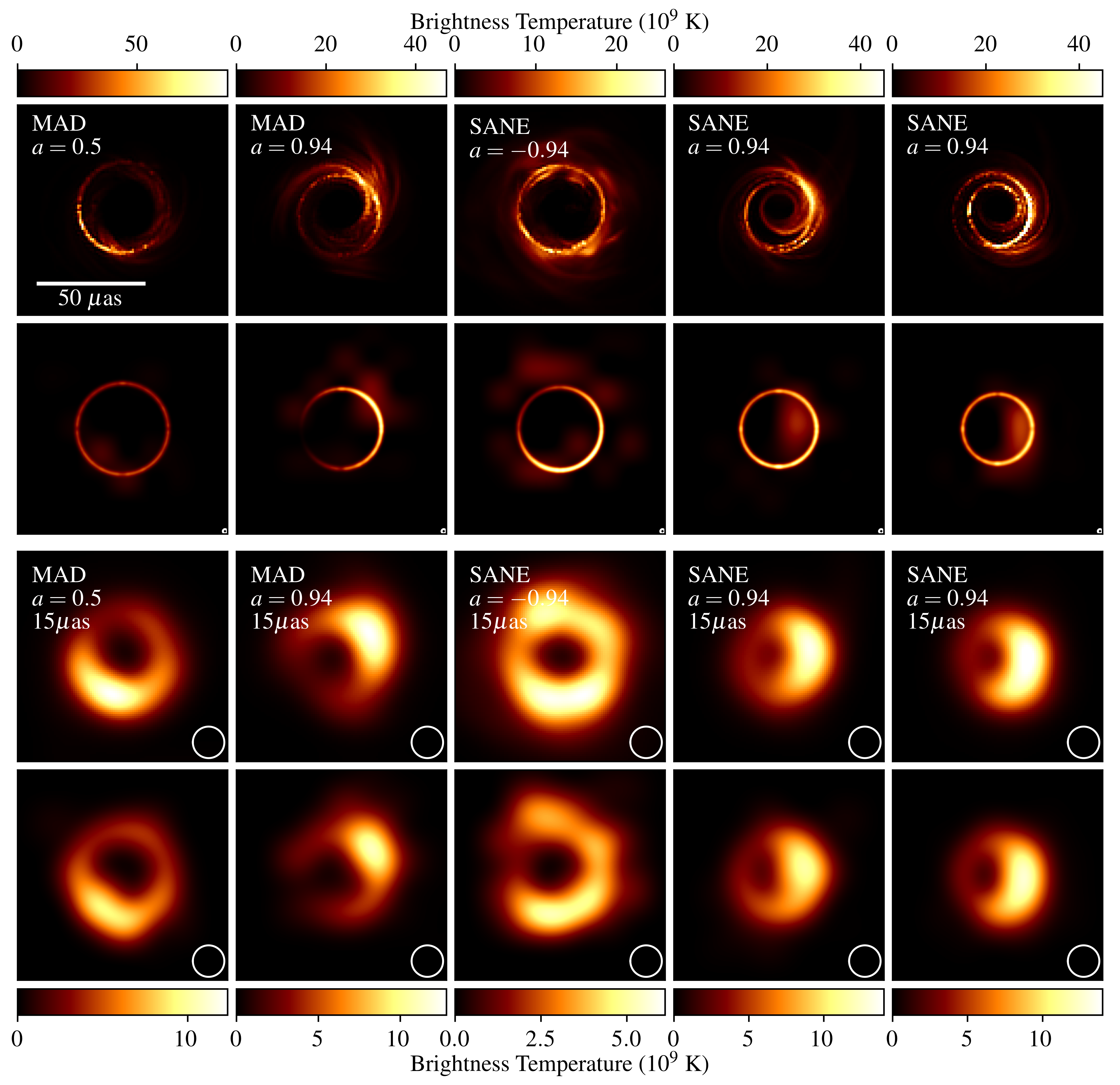}
  \end{center}
  \caption{Hybrid image reconstructions from simulated data generated from GRMHD simulation images appropriate for M87.  Each column corresponds to a single simulated data set, generated with the baseline coverage on April 10, 11, 5, 6 and 6 from left to right.  The top row shows the underlying truth image.  The second row shows the hybrid image comprised of a bicubic raster model with a slashed ring, smoothed with a Gaussian kernel with FWHM 2~$\mu$as.  The third and fourth rows show the truth and reconstructed hybrid images convolved with a Gaussian kernel with FWHM 15~$\mu$as, shown in the lower right corner.  For each column, the top two and bottom two rows have identical dynamic ranges, respectively, indicated by the associated color bars. The two rightmost columns are from the same GRMHD simulation taken from significantly different times ($>20$~days).}
  \label{fig:Jppcomp}
\end{figure*}

Figure~\ref{fig:Jppcomp} shows the reconstructions in comparison to the truths both before and after smoothing.  In each instance high-quality fits exist with $\chi^2/{\rm DoF}$ of 
$130/165$, $447/453$, $276/253$, $618/592$, and $539/596$, for the models as shown from left to right\footnote{The large variation in the number of degrees of freedom is due to the variety in the number of measurements on the different days of the EHT2017 observation campaign, with April 10 containing fewest measurements.}.

The slashed ring model component accurately reproduces the bright ringlike features in each image.  The remaining image component reconstructs the low-brightness extended emission.  There is little traction on the ring thickness, for which the posterior distribution remains dominated by the imposed uniform prior restricting the fractional width to below 5\%.  The weakness of this constraint is consistent with the fundamental resolution limits imposed by the $u$-$v$ coverage.  In contrast, the total flux and ring diameter are well constrained.  The combined ring component and image component together do an excellent job of reproducing the image morphology.  In all cases the ring component is modestly subdominant, contributing between 40\% and 60\% of the total flux in the image.

\begin{deluxetable*}{cccccccccc}
  \caption{Properties of GRMHD simulations in \autoref{fig:Jppcomp} (from left to right) and associated reconstructions}\label{tab:grmhdsims}
  \tablehead{
    \colhead{\hspace{0.125in}$a$}\hspace{0.125in} &
    \colhead{\hspace{0.125in}$i$}\hspace{0.125in} &
    \colhead{\hspace{0.125in}$R_{\rm hi}$}\hspace{0.125in} &
    \colhead{\hspace{0.125in}Flow}\hspace{0.125in} &
    \colhead{\hspace{0.125in}$r_{\rm ph}$}\hspace{0.125in} &
    \colhead{\hspace{0.125in}$r_{\rm peak}$}\hspace{0.125in} &
    \colhead{\hspace{0.125in}$r_{\rm ring}$}\hspace{0.125in} &
    \colhead{\hspace{0.125in}$F_{\rm total}$}\hspace{0.125in} &
    \colhead{\hspace{0.125in}$F_{\rm ring}/F_{\rm total}$}\hspace{0.125in} &
    \colhead{\hspace{0.125in}$w/r_{\rm ring}$}\hspace{0.125in}
    \vspace{-0.100in}\\
    & & &
    \colhead{Type} &
    \colhead{($\mu$as)} &
    \colhead{($\mu$as)} &
    \colhead{($\mu$as)} &
    \colhead{(Jy)} & &
  }
  \startdata
  $0.5$  & $158^\circ$ & $10$  & MAD  & 21.2 & 21.4 & $21.8_{-0.4}^{+0.3}$ & $0.50\pm0.03$ & $0.57_{-0.03}^{+0.04}$ & $0.02\pm0.01$ \\
  $0.94$ & $158^\circ$ & $40$  & MAD  & 18.7 & 19.3 & $19.3\pm1.0$         & $0.42\pm0.02$ & $0.44\pm0.05$          & $0.02_{-0.01}^{+0.02}$ \\
  $-0.94$ & $17^\circ$ & $20$  & SANE & 19.6 & 20.5 & $20.0\pm0.9$         & $0.37\pm0.02$ & $0.42\pm0.04$          & $0.03\pm0.02$ \\
  $0.94$ & $163^\circ$ & $160$ & SANE & 16.9 & 17.0 & $17.9_{-0.4}^{+0.3}$ & $0.47\pm0.02$ & $0.52_{-0.05}^{+0.04}$ & $0.03\pm0.02$ \\
  $0.94$ & $163^\circ$ & $160$ & SANE & 16.9 & 16.6 & $16.6_{-0.6}^{+0.5}$ & $0.47_{-0.01}^{+0.02}$ & $0.40\pm0.03$ & $0.03\pm0.02$ \\
  \enddata
  \tablecomments{Listed are the simulation parameters (spin, inclination, $R_{\rm hi}$, flow type) and the associated average asymptotic photon ring radius ($r_{\rm ph}$), radius of the peak of the azimuthally averaged flux ($r_{\rm peak}$), and reconstructed ring radius (median and 68\% confidence interval), total compact flux ($F_{\rm total}$), fraction of the flux contained in the ring ($F_{\rm ring}/F_{\rm total}$) and the fractional width of the ring ($w/r_{\rm ring}$).}
\end{deluxetable*}

The quantitative success of the ring reconstruction is presented in \autoref{tab:grmhdsims} and \autoref{fig:M87profiles}, where the azimuthally averaged intensity profiles from the ground truth image are compared against the posteriors on the ring radii.  In each case, the posterior covers in part or in its entirety the bright excess associated with the $n=1$ photon ring.  In the case of the magnetically arrested disk (MAD) $a=0.94$ model the posterior is double peaked, associated with the two apparent ringlike structures in the underlying truth image.

The fourth simulation (standard and normal evolution or SANE, $a=0.94$) has multiple ringlike features appearing at significantly different radii.  Of these, only one is associated with the gravitational, $n=1$ photon ring feature, the others with orbiting, turbulent features.  This results in multiple peaks in the azimuthally averaged brightness profile, limiting the formal applicability of the single-ring model.  Despite this, the ring radius posterior has a significant overlap with the brightness peak associated with the $n=1$ photon ring. 

However, these additional ringlike features are highly dynamic and expected to evolve significantly on timescales of weeks in M87. To determine how the ring reconstruction depends on the dynamical state of the emitting gas, we repeated the reconstruction exercise with a frame taken from the same GRMHD simulation taken at a sufficiently different time so as to be effectively uncorrelated.  This time, the resulting radius posterior matches the location of the bright peak with a similar fidelity to the other GRMHD simulation experiments. More importantly, the two ring radii reconstructions are inconsistent with each other at the 2.7$\sigma$ level. This suggests that repeated measurements separated by many dynamical times presents a natural strategy for distinguishing transient turbulent structures that arise due to astrophysical processes from the anticipated bright lensing features caused by gravity.

\subsection{Gravitational Inferences from Ring Reconstructions}
Encoded within the size and shape of the asymptotic photon ring is the structure of the underlying spacetime \citep{JP2010}. However, the ability to infer the gravitational properties of the black hole, e.g., the mass or spin, depends on the relationship between the location of the bright ringlike feature, presumably the $n=1$ photon ring, and the asymptotic photon ring. In all of the GRMHD simulations listed in \autoref{tab:grmhdsims} the peak of the bright ring is biased away from the asymptotic photon ring.  Three factors are relevant to the interpretation and magnitude of this bias.  

\begin{figure*}
    \begin{center}
        \includegraphics[width=\textwidth]{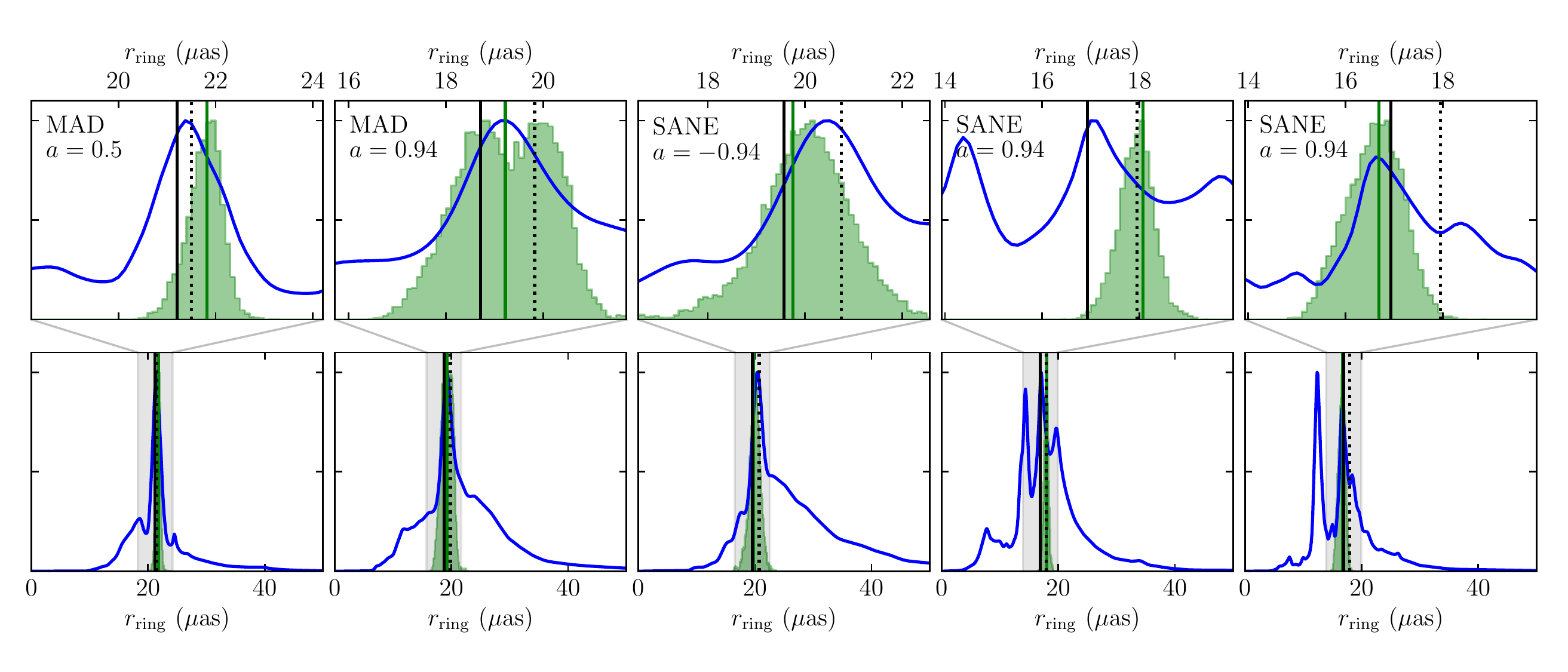}
    \end{center}
    \caption{Reconstructed ring sizes from simulated data generated from GRMHD simulations appropriate for M87.  Azimuthally averaged truth intensity profiles and the ring radius posteriors are shown in blue and green, respectively.  For reference the location of the photon ring for the GRMHD simulation is shown in black.  The center is chosen to coincide with that of the asymptotic photon ring.  The top panels show a zoom-in on the $\pm3~\mu$as region about the anticipated asymptotic photon ring diameter. In all panels, the most likely ring radius is shown by the green solid vertical line.  For comparison, the solid and dotted vertical black lines are the average photon ring radius and the $a=0$ asymptotic photon ring radius, respectively. The two rightmost columns are from the same GRMHD simulation taken from significantly different times ($>20$~days).
    } \label{fig:M87profiles}
\end{figure*}

First, note that the underlying resolution of the images from which the simulated data were constructed is $1~\mu$as.  This dominates the widths of the flux profiles in \autoref{fig:M87profiles}. Thus, even in principle, the precision with which the bright ring radius can be constrained is $\gtrsim0.1~\mu$as, limited fundamentally by the number of pixels covered by the ring.  In all instances in \autoref{tab:grmhdsims} the disparity is larger than this ostensible precision limit, though less than the underlying pixel resolution.  The role of GRMHD simulation resolution will be explored elsewhere. 

\begin{figure}[th!]
    \begin{center}
    \includegraphics[width=\columnwidth]{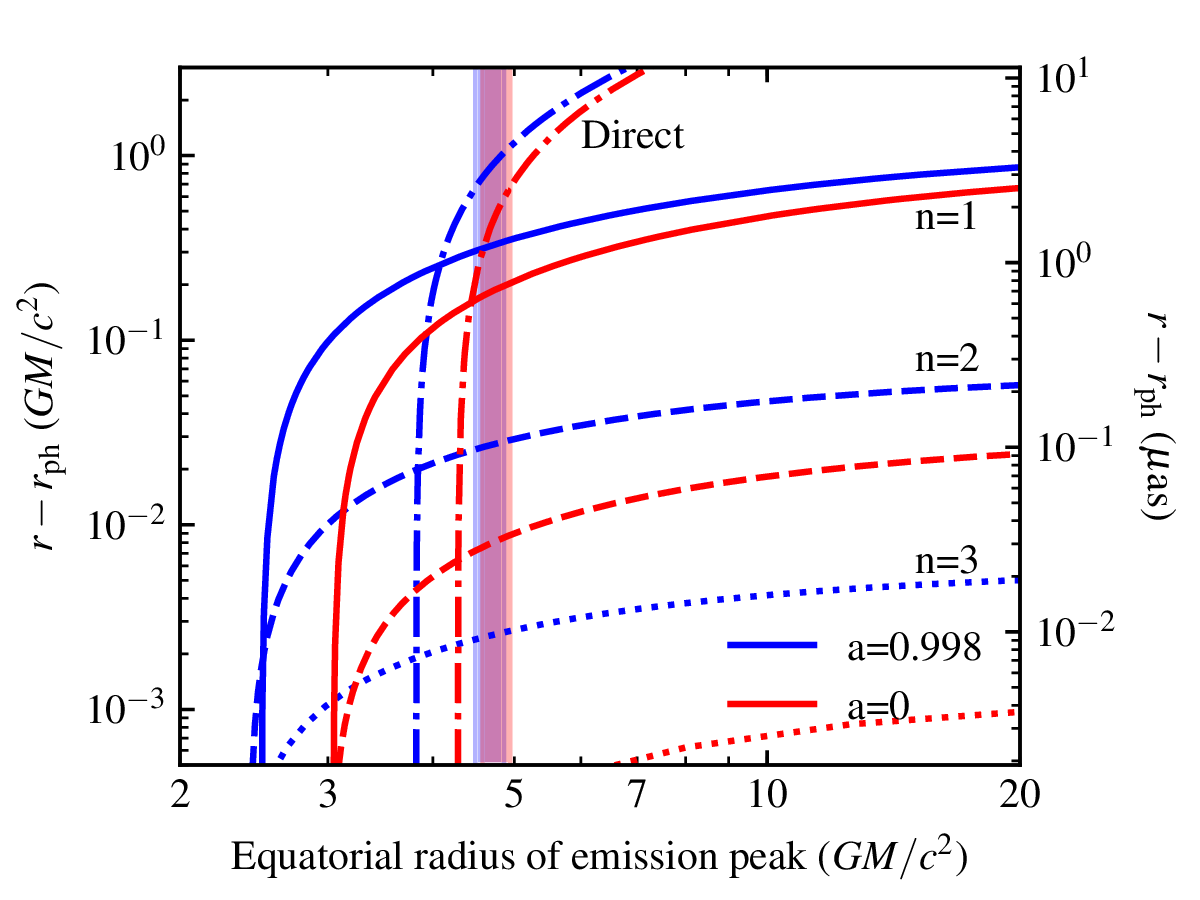}
    \end{center}
    \caption{Shift in the apparent radius of the brightness peak on a distant screen of the $n=1$, $n=2$, and $n=3$ photon rings relative to the asymptotic photon ring radius ($n\to\infty$; denoted by $r_{\rm ph}$) for emission within the equatorial plane viewed from the polar axis.  This shift is shown as a function of the equatorial Boyer-Lindquist radius of the peak emission, which maps directly onto the apparent location of the peak intensity within the appropriate photon ring.  The direct image component, formed by photons that deflect by less than $180^\circ$, is shown for reference.  The typical range of locations of the emission peak inferred for M87 from geometric crescent fits reported in \citetalias{M87_PaperVI} are indicated by the red and blue shaded regions.  These are comparable to the typical upper limits from GRMHD simulations.  The angular size of the shift for M87 is presented on the right side and assumes a fiducial angular size of the gravitational radius of $3.8~\mu$as.}\label{fig:deflection}
\end{figure}

Second, a dominant contribution to the ring flux is from the $n=1$ photon ring, which is produced by photons that have executed half orbits from emission to detection \citep[see, e.g.,][]{Johnson_2019}.  These are biased away from the asymptotic photon ring ($n\rightarrow\infty$).  Typical values for this bias from GRMHD simulations are $\approx 0.5~\mu$as (0.13~$GM/c^2D$) \citep{Johnson_2019} and are consistent with the biases observed in the first three GRMHD simulations in \autoref{tab:grmhdsims}.  The direction and precise value of this bias does depend on the distribution of emission near the black hole.  However a strong upper limit is imposed by the condition that polar photon trajectories deflect around the black hole sufficiently to return to the equatorial plane (see \autoref{fig:deflection} and \autoref{app:pri}), implying that the bias is less than $4.2~\mu$as (1.1~$GM/c^2D$) generally and typically considerably less than $1~\mu$as (0.26~$GM/c^2D$) for emission regions similar to those inferred for M87 from the geometric crescent fits described in \citetalias{M87_PaperVI} and found in GRMHD simulations of M87.

Third, dynamical features will present transient flux excesses inside and outside the asymptotic ring that bias the radial location of the flux peak.  This bias is clearly exemplified in the last two entries of \autoref{tab:grmhdsims}, where the radius of the flux peak within a single simulations varies by $0.4~\mu$as over long times ($>20$~days).  

However, despite the potential sources of bias in the location of flux peak, the posteriors of the reconstructed radii for the ring model component are consistent with the expected asymptotic photon ring radius.  In the first three entries of the \autoref{tab:grmhdsims}, for which a ringlike structure is present, the ring radius estimates are consistent with the asymptotic photon ring size at the 1$\sigma$ level.  For the final GRMHD simulation, the reconstructed ring radius estimates are consistent at the 2$\sigma$ level despite the presence of confounding features. An exhaustive study of the accuracy of the ring reconstructions for different underlying emission region morphologies will be presented in future publications.

\section{Conclusions}
\label{sec:conclusions}

Fully Bayesian image reconstruction from VLBI observations is now practical.  This method has been implemented in a fashion appropriate for EHT applications within the \themis analysis framework by treating the image reconstruction process identically with existing model-fitting tasks.  Employing \themis for this has exploited two key advantages: first, it permits the immediate and efficient application of high-performance computing resources to the reconstruction of radio images with realistic observational parameters and complications; the \themis image reconstruction tools can efficiently make use of thousands of cores simultaneously.  Second, it ensures that additional development within \themis is immediately available, including the implementation of additional samplers and mitigation of new sources of systematic error.  These image reconstruction tools have been successfully applied to a number of simulated EHT data sets.  Thus, Bayesian image reconstruction is now a practical tool for radio image analysis, though additional work on samplers may be required to fully realize its potential.

Performing image reconstruction in a Bayesian fashion presents two unique opportunities that arise due to treating the image reconstruction problem as a parameter estimation exercise in a statistically rigorous manner.  First, a consequence of this approach is the construction of a full image posterior, i.e., not a single best-fit image but rather a statistically meaningful collection of all of the images that are consistent with the data.  In parameter estimation applications, this posterior is the primary product.  In this case, it permits a quantitative assessment of the sources and magnitudes of errors.  For example, two-point functions on the image posterior elucidate the origin and size of the error induced by known holes in the $u$-$v$ coverage of the 2017 April EHT observations of M87.  More complex $n$-point functions may be developed in the future to address specific questions about image uniqueness and fidelity.

Second, treating image reconstruction and model parameter estimation identically enables seamless integration of the two methods of data analysis, a procedure we term hybrid imaging.  Importantly, the reconstructed posteriors on model parameter estimates are meaningful in the normal way.  

We also take this opportunity to demonstrate the ability to make a data-informed choice of the \fov and resolution, removing a frequent uncertainty in the image reconstruction process and source of subjective operator input.

We demonstrate hybrid imaging with two explicit applications.  The first is the inclusion of large-scale Gaussian components to model structure on scales much larger than the ostensible \fov, a procedure that has been utilized in previous EHT analyses \citepalias{M87_PaperIV,M87_PaperVI}.  The second and more exciting application is the inclusion of narrow ring model components to extract and characterize the $n=1$ photon rings from EHT observations of M87.  

Hybrid imaging of GRMHD simulations with realistic simulated EHT observations of M87 are able to faithfully reconstruct the properties of the prominent photon ring feature in the underlying truth images. These produce ring size posteriors that typically cover the true value within accuracies of 2\%--5\%.  Where turbulent features are mistaken for photon rings, the ring size estimates are variable on timescales of weeks in M87.  Thus, the benefits of non-parametric reconstructions combined with the strong restrictions on potential additional image features are already poised to significantly improve EHT constraints on the properties of the supermassive black hole in M87.

Photon ring characterization in Sagittarius A*, the supermassive black hole in the Galactic center, brings the additional challenges of addressing the interstellar scattering screen and short-timescale source variability. Both of these will be addressed in a future publication.  One considerable benefit of doing so is that high-precision mass and distance estimates exist for Sgr A* \citep{GRAVITY2019}, enabling precision tests of general relativity using the photon ring. 

There are many potential future applications of hybrid imaging.  All of these exploit its ability to constrain the freedom of image components via model specification, where such constraints are required due to experimental considerations (e.g., baseline coverage) or desirable because of high sensitivity.  Such applications within the context of the EHT include the connection of small-scale features with large-scale model structures, e.g., the modeling of the millimeter-wavelength milliarcsecond jet in M87 while imaging the horizon-scale millimeter-wavelength image structure. As the EHT adds stations and increases in sensitivity, hybrid imaging may be able to constrain the shape of the photon ring, producing an independent measure of black hole spin, and to detect higher-order rings, verifying a key predicted consequence of gravitational lensing. At the same time, hybrid imaging effectively separates the otherwise unresolved ring component from the larger-scale diffuse emission, and thus will enable dynamical studies on timescales as short as a week in M87. 

Finally, hybrid imaging permits an extension of the class of sources for which shadows may be detected and characterized beyond the two primary targets of the EHT, for which the shadow is formally resolvable.  By assuming that a crescent-like structure exists and contributes substantially to the emission in the bright radio cores, hybrid imaging extends the ability to estimate black hole masses to thousands of systems.

\mbox{}\\
\indent We thank Mareki Honma, Michael Johnson, Maciek Wielgus, and Katherine Boumann for useful comments. This work was made possible by the facilities of the Shared Hierarchical Academic Research Computing Network (SHARCNET:www.sharcnet.ca) and Compute/Calcul Canada (www.computecanada.ca).
Computations were made on the supercomputer Mammouth Parall\`ele 2 from University of Sherbrooke, managed by Calcul Qu\'ebec and Compute Canada. The operation of this supercomputer is funded by the Canada Foundation for Innovation (CFI), the minist\`ere de l'\'Economie, de la science et de l'innovation du Qu\'ebec (MESI) and the Fonds de recherche du Qu\'ebec - Nature et technologies (FRQ-NT).
This work was supported in part by Perimeter Institute for Theoretical Physics.  Research at Perimeter Institute is supported by the Government of Canada through the Department of Innovation, Science and Economic Development Canada and by the Province of Ontario through the Ministry of Economic Development, Job Creation and Trade.
A.E.B.\ thanks the Delaney Family for their generous financial support via the Delaney Family John A. Wheeler Chair at Perimeter Institute.
A.E.B.\ and P.T.\ receive additional financial support from the Natural Sciences and Engineering Research Council of Canada through a Discovery Grant. R.G.\ receives additional support from the ERC synergy grant “BlackHoleCam: Imaging the Event Horizon of Black Holes” (Grant No. 610058). 
D.W.P.\ was supported in part by the Black Hole Initiative at Harvard University, which is funded by grants from the John Templeton Foundation and the Gordon and Betty Moore Foundation to Harvard University.

\appendix

\section{Fourier domain cubic interpolation} \label{app:CubicInterpolation}
Here we derive the Fourier space representation of the bicubic convolution kernel given in Equation (\ref{eq:bcifour}).
The weight function in Equation (\ref{eq:bciweight}) is $C_2$.  However it does have discontinuities in the second and third derivatives, being identically zero beyond that.  Thus it may be represented as a triple and quadruple integral over a  small number of Dirac $\delta$ functions.

The second derivative of the cubic interpolation kernel is
\begin{equation}
  w''(x) =
  \begin{cases}
    0 & -2\ge x\\
    b \left[ -6x - 10 \right] & -1\ge x>-2\\
    -6(b+2) x - 2(b+3) & 0\ge x>1\\
    6(b+2) x - 2(b+3) & 1\ge x>0\\
    b \left[ 6x - 10 \right] & 2\ge x>1\\
    0 & x\ge 2    
  \end{cases}
\end{equation}
Note that this is not continuous, with the discontinuities of $\pm(-8b+6)$ and $\mp2b$ at $x=\pm1$ and $x=\pm2$, respectively.

The third derivative has two components, the first describing the smooth pieces of $w''(x)$ and the second consisting of $\delta$ functions associated with the discontinuities.  That is,
\begin{equation}
  w'''(x)
  =
  g(x) + F_3(x)
\end{equation}
where
\begin{equation}
  g(x)
  =
  \begin{cases}
    0 & -2\ge x\\
    -6b & -1\ge x>-2\\
    -6(b+2) & 0\ge x>-1\\
    6(b+2) & 1\ge x>0\\
    6b & 2>\ge x>1\\
    0 & x\ge 2,
  \end{cases}
\end{equation}
and
\begin{equation}
  \begin{aligned}
  F_3(x)
  =
  & 2b\delta(x+2)
  +(8b+6)\delta(x+1) \\
  & -(8b+6)\delta(x-1)
  -2b\delta(x-2).
  \end{aligned}
\end{equation}
Again, there are discontinuities in the non-singular part of $w'''$, i.e., $g(x)$, though now at $x=0$, $x=\pm1$, and $x=\pm2$; $g(x)$ is constant otherwise.  Thus, the fourth derivative is consists solely of $\delta$ functions, $w^{(iv)}(x)=F_4(x)$ where
\begin{equation}
  \begin{aligned}
  F_4(x) = &
  -6b\delta(x+2)
  -12\delta(x+1)\\
  & +12(b+2)\delta(x)
  -12\delta(x-1)
  -6b\delta(x-2).
  \end{aligned}
\end{equation}
Then, noting that all constants of integration vanish (because $f(x)=0$ for $|x|>2$), we can write
\begin{equation}
  w(x) = \iiint F_3(x) + \iiiint F_4(x).
\end{equation}

The frequency-space representation of $F_3$ and $F_4$ are:
\begin{equation}
  \begin{aligned}
    F_3(k)
    &= \int dx e^{-ikx} F_3(x)\\
    &= 4i \sin(k) \left[ 2b\cos(k) + (4b+3) \right]\\
    F_4(k)
    &= \int dx e^{-ikx} F_4(x)\\
    &= 12b\left[1-\cos(2k)\right] + 24\left[1-\cos(k)\right],
  \end{aligned}
\end{equation}
where $k=2\pi\Delta u$, where $\Delta$ is the spacing between control points.
Therefore, applying the integrals,
\begin{equation}
  \begin{aligned}
    W(k)
    &= \frac{F_3(k)}{(ik)^3} + \frac{F_4(k)}{(ik)^4}\\
    &= - \frac{4}{k^3} \sin(k) \left[2b\cos(k)+(4b+3)\right]\\
    & \quad + \frac{12}{k^4} \left\{ b \left[ 1-\cos(2k) \right] + 2\left[1-\cos(k)\right] \right\}.
  \end{aligned}
\end{equation}

\section{Photon ring images} \label{app:pri}
In \autoref{fig:deflection} we show the mapping between points on the equatorial and image planes for observers on the polar axis.  Here we collect information about how this mapping is generated practically and what it implies generally.

The mapping is generated by direct integration of the null geodesic equations for photons leaving a distant screen.  We simplify the problem by considering an observer on the polar axis, for which axisymmetry reduces the function to a one-dimensional $\Delta r_n(R)$, which is a mapping of the Boyer-Lindquist radius $R$ in the equatorial plane (at the $(n+1)$th photon crossing) to the radial location in the image plane. While this does not generalize to arbitrary inclinations, the insensitivity of the asymptotic photon ring radius to inclinations below $30^\circ$ suggests that this is sufficient to estimate the magnitudes of the shifts of the low-order photon ring radii from the asymptotic value \citep{JP2010}.

This mapping is monotonic, a fact that has two important consequences.  First, the entirety of the equatorial plane is remapped at every order, and thus the photon rings present a sequence of copies of the direct emission.  Second, and more relevant here, is that the peak of the intensity within the $n$th photon ring is fully determined by the location peak of the emission, $R_{\rm pk}$ and the map.  This arises immediately from the conservation of $I_\nu/\nu^3$ along the photon trajectories.  Thus, for optically thin emission within the equatorial plane, the total intensity at a radial position of $r$ in the image plane is
\begin{equation}
\frac{I_\nu}{\nu^3}(r)
=
\sum_{m=0}^{n} \frac{I_{g\nu}}{(g\nu)^3}(R_m),
\end{equation}
where $g$ is the standard Doppler factor, $\Delta r_m(R_m) \equiv r - r_{\rm ph}$ defines $R_m$, and $n$ is the highest-order image present at $r$.  This is peaked when
\begin{equation}
  \begin{aligned}
    \frac{d(I^{\rm obs}_\nu/\nu^3)}{d\Delta r}
    &=
    \sum_{m=0}^{n}\frac{d[I^{\rm em}_{g\nu}/(g\nu)^3]}{d R_m} \frac{d R_m}{d\Delta r}\\
    &\approx
    \frac{d[I^{\rm em}_{g\nu}/(g\nu)^3]}{d R_n} \frac{d R_n}{d\Delta r} = 0,
  \end{aligned}
\end{equation}
where we have used the fact that the ring widths decrease exponentially with $n$, and thus for smoothly varying emission regions $dR_n/d\Delta r$ increases exponentially with $n$ \citep[see, e.g.,][]{Johnson_2019}.  Therefore, if $R_{\rm pk}$ is the location of the peak of $I^{\rm em}_{g\nu}/(g\nu)^3$, i.e., the peak of the emission after inclusion of the Doppler shift and Doppler beaming, then the peak of the $n$th-order photon ring emission is $\Delta r_{n,\rm pk} \approx \Delta r_n(R_{\rm pk})$.


\mbox{}\\

\bibliographystyle{aasjournal_aeb}
\bibliography{references}

\begin{thebibliography}{}
\expandafter\ifx\csname natexlab\endcsname\relax\def\natexlab#1{#1}\fi
\providecommand{\url}[1]{\href{#1}{#1}}
\providecommand{\dodoi}[1]{doi:~\href{http://doi.org/#1}{\nolinkurl{#1}}}
\providecommand{\doeprint}[1]{\href{http://ascl.net/#1}{\nolinkurl{http://ascl.net/#1}}}
\providecommand{\doarXiv}[1]{\href{https://arxiv.org/abs/#1}{\nolinkurl{https://arxiv.org/abs/#1}}}

\bibitem[{{Akiyama} {et~al.}(2017{\natexlab{a}}){Akiyama}, {Kuramochi},
  {Ikeda}, {Fish}, {Tazaki}, {Honma}, {Doeleman}, {Broderick}, {Dexter},
  {Mo{\'s}cibrodzka}, {Bouman}, {Chael}, \& {Zaizen}}]{Akiyama_2017a}
{Akiyama}, K., {Kuramochi}, K., {Ikeda}, S., {et~al.} 2017{\natexlab{a}}, \apj,
  838, 1

\bibitem[{{Akiyama} {et~al.}(2017{\natexlab{b}}){Akiyama}, {Ikeda}, {Pleau},
  {Fish}, {Tazaki}, {Kuramochi}, {Broderick}, {Dexter}, {Mo{\'s}cibrodzka},
  {Gowanlock}, {Honma}, \& {Doeleman}}]{Akiyama_2017b}
{Akiyama}, K., {Ikeda}, S., {Pleau}, M., {et~al.} 2017{\natexlab{b}}, \aj, 153,
  159

\bibitem[{{Arras} {et~al.}(2019){Arras}, {Frank}, {Leike}, {Westermann}, \&
  {En{\ss}lin}}]{Arras_2019}
{Arras}, P., {Frank}, P., {Leike}, R., {Westermann}, R., \& {En{\ss}lin}, T.~A.
  2019, \aap, 627, A134

\bibitem[{{Baldwin} \& {Warner}(1978)}]{Baldwin_1978}
{Baldwin}, J.~E., \& {Warner}, P.~J. 1978, \mnras, 182, 411

\bibitem[{{Blackburn} {et~al.}(2020){Blackburn}, {Pesce}, {Johnson}, {Wielgus},
  {Chael}, {Christian}, \& {Doeleman}}]{Blackburn+2019}
{Blackburn}, L., {Pesce}, D.~W., {Johnson}, M.~D., {et~al.} 2020, \apj, 894, 31

\bibitem[{{Broderick} {et~al.}(2020){Broderick}, {Gold}, {Karami},
  {Preciado-L{\'o}pez}, {Tiede}, {Pu}, {Akiyama}, {Alberdi}, {Alef}, {Asada},
  {Azulay}, {Baczko}, {Balokovi{\'c}}, {Barrett}, {Bintley}, {Blackburn},
  {Boland}, {Bouman}, {Bower}, {Bremer}, {Brinkerink}, {Brissenden}, {Britzen},
  {Broguiere}, {Bronzwaer}, {Byun}, {Carlstrom}, {Chael}, {Chatterjee},
  {Chatterjee}, {Chen}, {Chen}, {Cho}, {Conway}, {Cordes}, {Crew}, {Cui},
  {Davelaar}, {De Laurentis}, {Deane}, {Dempsey}, {Desvignes}, {Doeleman},
  {Eatough}, {Falcke}, {Fish}, {Fomalont}, {Fraga-Encinas}, {Friberg}, {Fromm},
  {Galison}, {Gammie}, {Garc{\'\i}a}, {Gentaz}, {Georgiev}, {Goddi},
  {G{\'o}mez}, {Gu}, {Gurwell}, {Hada}, {Hecht}, {Hesper}, {Ho}, {Ho}, {Honma},
  {Huang}, {Huang}, {Hughes}, {Inoue}, {Issaoun}, {James}, {Janssen}, {Jeter},
  {Jiang}, {Jim{\'e}nez-Rosales}, {Johnson}, {Jorstad}, {Jung}, {Karuppusamy},
  {Kawashima}, {Keating}, {Kettenis}, {Kim}, {Kim}, {Kino}, {Koay}, {Koch},
  {Koyama}, {Kramer}, {Kramer}, {Krichbaum}, {Kuo}, {Lee}, {Li}, {Li},
  {Lindqvist}, {Lico}, {Liu}, {Liuzzo}, {Lo}, {Lobanov}, {Loinard}, {Lonsdale},
  {Lu}, {MacDonald}, {Mao}, {Marscher}, {Mart{\'\i}-Vidal}, {Matsushita},
  {Matthews}, {Menten}, {Mizuno}, {Mizuno}, {Moran}, {Moriyama},
  {Moscibrodzka}, {M{\"u}ller}, {Nagai}, {Nagar}, {Nakamura}, {Narayan},
  {Narayanan}, {Natarajan}, {Neri}, {Ni}, {Noutsos}, {Okino}, {Olivares},
  {Ortiz-Le{\'o}n}, {Oyama}, {Palumbo}, {Park}, {Pen}, {Pesce}, {Pi{\'e}tu},
  {Plambeck}, {PopStefanija}, {Porth}, {Prather}, {Ramakrishnan}, {Rao},
  {Rawlings}, {Raymond}, {Rezzolla}, {Ripperda}, {Roelofs}, {Rogers}, {Ros},
  {Rose}, {Rottmann}, {Ruszczyk}, {Ryan}, {Rygl}, {S{\'a}nchez},
  {S{\'a}nchez-Arguelles}, {Sasada}, {Savolainen}, {Schloerb}, {Schuster},
  {Shao}, {Shen}, {Small}, {Sohn}, {SooHoo}, {Tazaki}, {Tilanus}, {Titus},
  {Toma}, {Torne}, {Traianou}, {Trippe}, {Tsuda}, {van Bemmel}, {van
  Langevelde}, {van Rossum}, {Wagner}, {Wardle}, {Weintroub}, {Wex}, {Wharton},
  {Wielgus}, {Wong}, {Wu}, {Yoon}, {Young}, {Young}, {Younsi}, {Yuan}, {Yuan},
  {Zensus}, {Zhao}, {Zhao}, {Zhu}, \& {Event Horizon Telescope
  Collaboration}}]{themis}
{Broderick}, A.~E., {Gold}, R., {Karami}, M., {et~al.} 2020, \apj, 897, 139

\bibitem[{{Cai} {et~al.}(2018{\natexlab{a}}){Cai}, {Pereyra}, \&
  {McEwen}}]{Cai_2018a}
{Cai}, X., {Pereyra}, M., \& {McEwen}, J.~D. 2018{\natexlab{a}}, \mnras, 480,
  4154

\bibitem[{{Cai} {et~al.}(2018{\natexlab{b}}){Cai}, {Pereyra}, \&
  {McEwen}}]{Cai_2018b}
---. 2018{\natexlab{b}}, \mnras, 480, 4170

\bibitem[{{Chael} {et~al.}(2018){Chael}, {Johnson}, {Bouman}, {Blackburn},
  {Akiyama}, \& {Narayan}}]{Chael_2018}
{Chael}, A.~A., {Johnson}, M.~D., {Bouman}, K.~L., {et~al.} 2018, \apj, 857, 23

\bibitem[{{Chael} {et~al.}(2016){Chael}, {Johnson}, {Narayan}, {Doeleman},
  {Wardle}, \& {Bouman}}]{Chael_2016}
{Chael}, A.~A., {Johnson}, M.~D., {Narayan}, R., {et~al.} 2016, \apj, 829, 11

\bibitem[{{Clark}(1980)}]{Clark_1980}
{Clark}, B.~G. 1980, \aap, 89, 377

\bibitem[{{Event Horizon Telescope Collaboration}
  {et~al.}(2019{\natexlab{a}})}]{M87_PaperIII}
{Event Horizon Telescope Collaboration}, {et~al.} 2019{\natexlab{a}}, \apjl,
  875, L3

\bibitem[{{Event Horizon Telescope Collaboration}
  {et~al.}(2019{\natexlab{b}})}]{M87_PaperI}
---. 2019{\natexlab{b}}, \apjl, 875, L1

\bibitem[{{Event Horizon Telescope Collaboration}
  {et~al.}(2019{\natexlab{c}})}]{M87_PaperII}
---. 2019{\natexlab{c}}, \apjl, 875, L2

\bibitem[{{Event Horizon Telescope Collaboration}
  {et~al.}(2019{\natexlab{d}})}]{M87_PaperIV}
---. 2019{\natexlab{d}}, \apjl, 875, L4

\bibitem[{{Event Horizon Telescope Collaboration}
  {et~al.}(2019{\natexlab{e}})}]{M87_PaperV}
---. 2019{\natexlab{e}}, \apjl, 875, L5

\bibitem[{{Event Horizon Telescope Collaboration}
  {et~al.}(2019{\natexlab{f}})}]{M87_PaperVI}
---. 2019{\natexlab{f}}, \apjl, 875, L6

\bibitem[{{Frieden}(1972)}]{Frieden_1972}
{Frieden}, B.~R. 1972, Journal of the Optical Society of America (1917-1983),
  62, 511

\bibitem[{{Gravity Collaboration} {et~al.}(2019){Gravity Collaboration},
  {Abuter}, {Amorim}, {Baub{\"o}ck}, {Berger}, {Bonnet},
  {et~al.}}]{GRAVITY2019}
{Gravity Collaboration}, {Abuter}, R., {Amorim}, A., {et~al.} 2019, \aap, 625,
  L10

\bibitem[{{Greiner} {et~al.}(2016){Greiner}, {Vacca}, {Junklewitz}, \&
  {En{\ss}lin}}]{Greiner_2016}
{Greiner}, M., {Vacca}, V., {Junklewitz}, H., \& {En{\ss}lin}, T.~A. 2016,
  arXiv e-prints.
\newblock \doarXiv{1605.04317}

\bibitem[{{Gull} \& {Daniell}(1978)}]{Gull_1978}
{Gull}, S.~F., \& {Daniell}, G.~J. 1978, \nat, 272, 686

\bibitem[{Hilbert(1917)}]{Hilbert1917}
Hilbert, D. 1917, Weidmannsche Buchhandlung, Berlin, 53

\bibitem[{{H{\"o}gbom}(1974)}]{Hogbom_1974}
{H{\"o}gbom}, J.~A. 1974, \aaps, 15, 417

\bibitem[{{Honma} {et~al.}(2014){Honma}, {Akiyama}, {Uemura}, \&
  {Ikeda}}]{Honma_2014}
{Honma}, M., {Akiyama}, K., {Uemura}, M., \& {Ikeda}, S. 2014, \pasj, 66, 95

\bibitem[{{Huijser} {et~al.}(2015){Huijser}, {Goodman}, \&
  {Brewer}}]{Huijser+2015}
{Huijser}, D., {Goodman}, J., \& {Brewer}, B.~J. 2015, arXiv e-prints.
\newblock \doarXiv{1509.02230}

\bibitem[{{Johannsen} \& {Psaltis}(2010)}]{JP2010}
{Johannsen}, T., \& {Psaltis}, D. 2010, \apj, 718, 446

\bibitem[{{Johnson} {et~al.}(2020){Johnson}, {Lupsasca}, {Strominger}, {Wong},
  {Hadar}, {Kapec}, {Narayan}, {Chael}, {Gammie}, {Galison}, {Palumbo},
  {Doeleman}, {Blackburn}, {Wielgus}, {Pesce}, {Farah}, \&
  {Moran}}]{Johnson_2019}
{Johnson}, M.~D., {Lupsasca}, A., {Strominger}, A., {et~al.} 2020, Science
  Advances, 6, eaaz1310

\bibitem[{{Junklewitz} {et~al.}(2016){Junklewitz}, {Bell}, {Selig}, \&
  {En{\ss}lin}}]{Junklewitz_2016}
{Junklewitz}, H., {Bell}, M.~R., {Selig}, M., \& {En{\ss}lin}, T.~A. 2016,
  \aap, 586, A76

\bibitem[{{Kuramochi} {et~al.}(2018){Kuramochi}, {Akiyama}, {Ikeda}, {Tazaki},
  {Fish}, {Pu}, {Asada}, \& {Honma}}]{Kuramochi_2018}
{Kuramochi}, K., {Akiyama}, K., {Ikeda}, S., {et~al.} 2018, \apj, 858, 56

\bibitem[{{Lochner} {et~al.}(2015){Lochner}, {Natarajan}, {Zwart}, {Smirnov},
  {Bassett}, {Oozeer}, \& {Kunz}}]{Lochner_2015}
{Lochner}, M., {Natarajan}, I., {Zwart}, J. T.~L., {et~al.} 2015, \mnras, 450,
  1308

\bibitem[{{Naghibzadeh} \& {van der Veen}(2018)}]{Naghibzadeh_2018}
{Naghibzadeh}, S., \& {van der Veen}, A.-J. 2018, \mnras, 479, 5638

\bibitem[{{Narayan} \& {Nityananda}(1986)}]{Narayan_1986}
{Narayan}, R., \& {Nityananda}, R. 1986, \araa, 24, 127

\bibitem[{{Pearson} \& {Readhead}(1984)}]{Pearson_1984}
{Pearson}, T.~J., \& {Readhead}, A.~C.~S. 1984, \araa, 22, 97

\bibitem[{{Schwab}(1984)}]{Schwab_1984}
{Schwab}, F.~R. 1984, \aj, 89, 1076

\bibitem[{{Schwarz}(1978)}]{Schwarz_1978}
{Schwarz}, U.~J. 1978, \aap, 65, 345

\bibitem[{{Sutter} {et~al.}(2014){Sutter}, {Wandelt}, {McEwen}, {Bunn},
  {Karakci}, {Korotkov}, {Timbie}, {Tucker}, \& {Zhang}}]{Sutter_2014}
{Sutter}, P.~M., {Wandelt}, B.~D., {McEwen}, J.~D., {et~al.} 2014, \mnras, 438,
  768

\end{thebibliography}

\end{document}